\documentclass[doubleblind]{ecai} 

\usepackage{latexsym}
\usepackage{amssymb}
\usepackage{amsmath}
\usepackage{amsthm}
\usepackage{booktabs}
\usepackage{enumitem}
\usepackage{graphicx}
\usepackage{color}
\usepackage{soul}

\usepackage{wrapfig}

\newtheorem{theorem}{Theorem}

\newcommand{\BibTeX}{B\kern-.05em{\sc i\kern-.025em b}\kern-.08em\TeX}

\newif\ifhide
\hidetrue 

\begin{document}


\begin{frontmatter}


\paperid{2559} 


\title{B2MAPO: A Batch-by-Batch Multi-Agent Policy Optimization to Balance Performance and Efficiency}


\author[A]{\fnms{Wenjing}~\snm{ZHANG}\orcid{0009-0006-2571-604X}}
\author[A]{ \fnms{Wei}
~\snm{ZHANG} \orcid{0000-0002-0598-4606}\thanks{Corresponding Author. Email: weizhang@hit.edu.cn}}
\author[A]{\fnms{Wenqing }~\snm{HU} }
\author[A]{\fnms{Yifan}~\snm{WANG} }

\address[A]{Harbin Institute of Technology}


\begin{abstract}
Most multi-agent reinforcement learning approaches adopt two types of policy optimization methods that either update policy simultaneously or sequentially. 
Simultaneously updating policies of all agents introduces non-stationarity problem. 
Although sequentially updating policies agent-by-agent in an appropriate order improves policy performance, it is prone to low efficiency due to sequential execution, resulting in longer model training and execution time.
Intuitively, partitioning policies of all agents according to their interdependence and updating joint policy batch-by-batch can effectively balance performance and efficiency. 
However, how to determine the optimal batch partition of policies and batch updating order are challenging problems.
Firstly, a sequential batched policy updating scheme, B2MAPO (\textbf{B}atch by \textbf{B}atch \textbf{M}ulti-\textbf{A}gent \textbf{P}olicy \textbf{O}ptimization), is proposed with a theoretical guarantee of the monotonic incrementally tightened bound.
Secondly, a universal modulized plug-and-play B2MAPO hierarchical framework, which satisfies CTDE principle, is designed to conveniently integrate any MARL models to fully exploit and merge their merits, including policy optimality and inference efficiency.
Next, a DAG-based B2MAPO algorithm is devised, which is a carefully designed implementation of B2MAPO framework. 
The upper layer employs PPO algorithm with attention mechanism to reveal interdependence between policies, and generates DAGs of agents which are used to produce optimal batch sequence through topological sorting. 
The lower layer trains two joint policies in parallel and minimizes KL divergence between them periodically.
One joint policy is sequentially updated according to B2MAPO scheme with batch sequence, another derived joint policy is simultaneously updated by MAPPO. 
While decentralized execution, only the derived joint policy is adopted for decision-making. 
Comprehensive experimental results conducted on StarCraftII Multi-agent Challenge and Google Football Research demonstrate the performance of DAG-based B2MAPO algorithm outperforms baseline methods.
Meanwhile, compared with A2PO, our algorithm reduces the model training and execution time by 60.4\% and 78.7\% , respectively.
\end{abstract}

\end{frontmatter}


\section{Introduction}

\begin{figure}[ht]
\setlength{\abovecaptionskip}{-0.1cm}
\begin{center}
\centerline{\includegraphics[width=0.88\columnwidth]{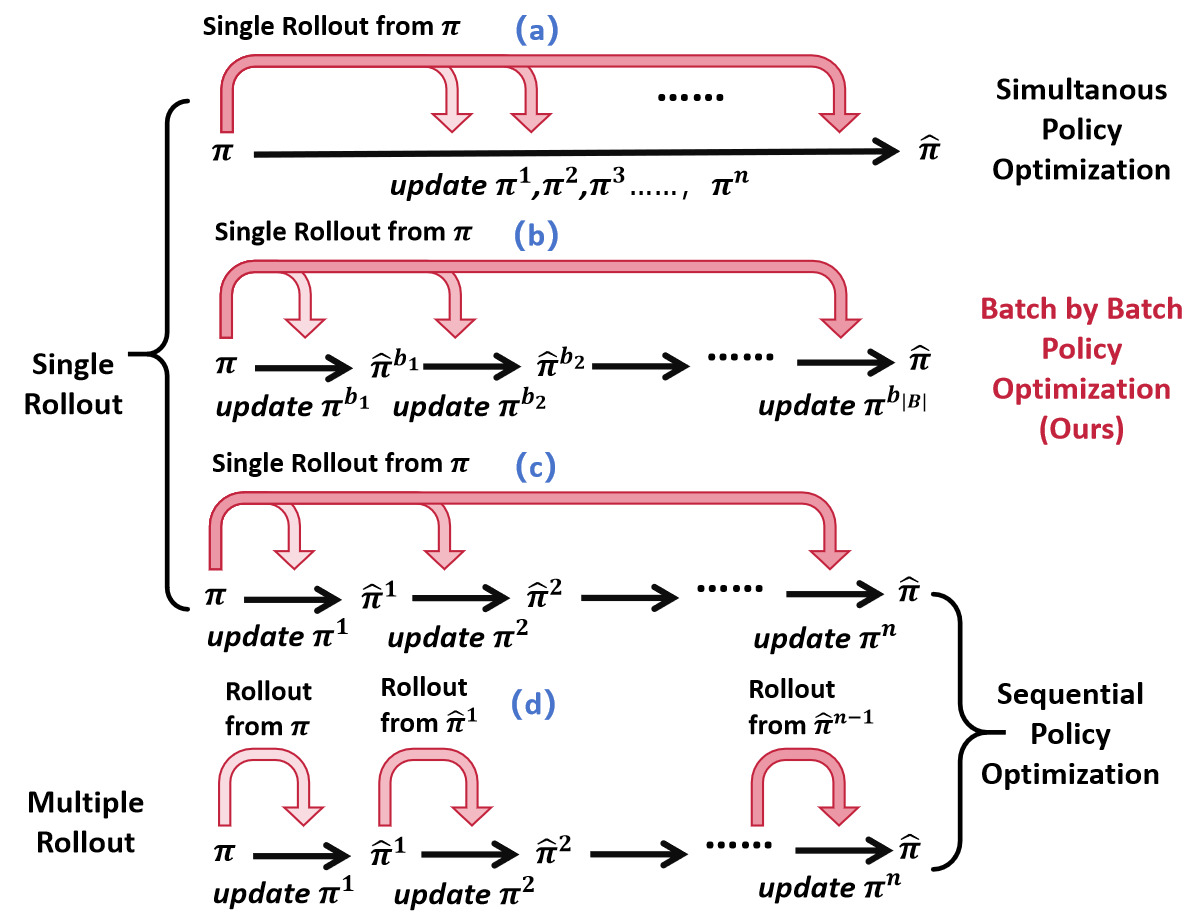}}
\caption{The taxonomy of different policy rollout and update schemes. 
The joint policy at the thick red arrow tail is used for rollout, and the sampled data is used to update the joint policy pointed by the arrow.
The thin black arrow indicates that the old joint policy at the arrow tail is updated to become the new joint policy pointed by the arrow.
For batch-by-batch policy optimization, $\boldsymbol{\pi}^{b_k}$ represents the joint policy consisting of the policies of multiple agents in batch ${b_k}$.$\hat{\boldsymbol{\pi}}^{b_k}$ means the updated joint policy after batch ${b_k}$ is updated. 
}
\label{rollout scheme}
\end{center}
\end{figure}

Cooperative Multi-Agent Reinforcement Learning (MARL) methods have addressed numerous challenges in both virtual and real-world scenarios, such as traffic signal control\cite{Liu_Luo_Yuan_Li_Jin_Chen_Pan, Jiang_Li_Sun_Zheng_2021}, automated freight handling\cite{2021Multi}, and autonomous driving\cite{2019Multi,2019Multi_2}. 
Cooperative MARL aims to train a cluster of agents to collaborate on tasks to maximize the expected cumulative rewards.
The existing MARL policy optimization methods are mainly divided into two categories: simultaneous policy optimization and sequential policy optimization. 
Simultaneous policy optimization (Figure \ref{rollout scheme}a) like MAPPO\cite{Yu_Velu_Vinitsky_Wang_Bayen_Wu_2021} typically assumes that agents' policies are independent. 
All agents update their policies simultaneously and cannot predict the dynamic policy changes of other agents. 
Therefore, due to dynamic policy improvements of other agents, from the perspective of one agent, the environment changes dynamically, which causes non-stationarity problems\cite{Hernandez-Leal_Kaisers_Baarslag_Cote_2017}.
Consequently, the non-stationarity problems directly result in sub-optimal performance in complex task scenarios. 
To address this, the Centralized Training with Decentralized Execution (CTDE) \cite{Lowe_Wu_Tamar_Harb_Abbeel_Mordatch_2017} paradigm has been proposed. 
During centralized training, agents can access to other agents' information and the global state, while during decentralized execution, agents execute decisions independently based on their individual policies. 
However, the non-stationarity problem still persists, especially when agents only have partial observations, making policy optimization more challenging.

Conversely, sequential policy optimization follows a sequential update scheme. 
The sequential update scheme allows the agent to perceive the dynamic changes produced by preceding agents' policies, providing another perspective for analyzing the interactive dependencies of agents\cite{Gemp_Chen_McWilliams_2022}. 
Sequential policy optimization includes multiple rollout and single rollout methods. 
RSIPA\cite{Bertsekas_2021}, the representative of multiple rollout methods(Figure \ref{rollout scheme}d), effectively solves non-stationary problems by converting non-stationary multi-agent reinforcement learning problems into stationary single-agent reinforcement learning (SARL) problems\cite{Kakade_Langford_2002}. 
Although RSIPA demonstrates robust performance, its sample efficiency and inference efficiency are extremely lower than that of MARL methods with simultaneous policy updating scheme.
Then, the single-rollout methods(Figure \ref{rollout scheme}c) are proposed to improve sample and inference efficiency. 
Heterogeneous-agent Proximal Policy Optimization (HAPPO) \cite{Kuba_Chen_Wen_Wen_Sun_Wang_Yang_2021} utilizes the single rollout samples to estimate local advantages for policy optimization, which improves sample efficiency. 
Although HAPPO introduces the first multi-agent trust region method that enjoys a theoretically justified monotonic improvement guarantee, it results in poor performance due to the lack of a theoretical guarantee for individual policy improvement.
Furthermore, HAPPO still has low efficiency in the training phase because of its sequential predetermined policy updating scheme.
Recently, Agent-by-agent policy optimization (A2PO)\cite{Wang_Tian_Wan_Wen_Wang_Zhang_2023} theoretically guarantees monotonic improvement of both joint and individual policies. 
However, due to the sequential update scheme, A2PO also exhibits lower efficiency during training. 
Moreover, the training time increases with the number of agents, leading to higher training and execution time for joint policy optimization in large-scale agent clusters. 
Due to errors in estimating the advantage function, A2PO may still fall into a sub-optimum.
Meanwhile, the sequential updates of HAPPO and A2PO are established upon the assumption that each agent’s update does not depend on and impact the policies of others.
However, this assumption, becomes inapplicable with parameter sharing, as each agent’s policy update can induce changes in the policy network parameters of other agents, thereby the optimality cannot be guaranteed in practice.
A2PO alleviates this problem by a heuristic approach, which preferentially updates agents with large advantage functions. 
However, relying solely on advantages to determine the update sequence may be overly simplistic, since it ignores the dynamic dependencies among agents. 

In order to improve the efficiency of the model training, independent policies can be updated simultaneously, while interdependent policies should be updated sequentially. 
To achieve this, we first introduce the concept of "\textbf{batch}", which divides the policies into partitions (groups), and then policy optimization can be performed in batch order rather than agent order. 
It is required to partition policies into batches and order the batches. 
In order to prevent the non-stationary problem, all agents in the same batch must be independent to update simultaneously. 
However, the policy dependency of agents is very hard to obtain in practice. 
Moreover, the minimum number of partitioned batches can be modeled as a set cover problem(SCP)\cite{caprara2000algorithms}, which is known to be NP-hard.

To ensure optimal performance, the surrogate objective based on offline policy correction is proposed, which theoretically guarantees the incrementally tightened monotonic bound with the monotonic improvement of both joint and individual policies. 
The theoretical properties lead to Batch by Batch Multi-Agent Policy Optimization(B2MAPO), a novel sequential batch updating scheme(Figure \ref{rollout scheme}d). 
Training efficiency of B2MAPO is higher than that of A2PO because it updates policies in batch order rather than agent order.

Based on B2MAPO scheme, a universal modulized plug-and-play B2MAPO hierarchical framework, which satisfies CTDE principle, is designed to conveniently integrate any MARL models to fully exploit and merge their merits, including policy optimality and inference efficiency. 
Meanwhile, according to theoretical proof that for any joint policy $\boldsymbol{\pi}$ trained by B2MAPO scheme, there always exists a joint policy $\boldsymbol{\pi}_{ind}$ produced by CTDE-based MARL algorithms with equivalent expected cumulative discounted rewards. 
In the lower-layer network design of the B2MAPO framework, two joint policies are trained in parallel and minimizes KL divergence between them periodically. 
One joint policy is sequentially updated according to B2MAPO scheme with batch sequence, another derived joint policy is simultaneously updated by MAPPO. 
The derived joint policy is trained through diverse experiences collected by joint policy acquired by B2MAPO scheme. 
Finally, during the execution phase, only the derived joint policy is used for decision-making to improve execution efficiency. 
Ultimately, the derived joint policy achieves high execution efficiency comparable to the CTDE-based MARL algorithms while ensuring algorithm optimality.

Finally, a carefully designed implementation of B2MAPO framework is devised, called the DAG-based B2MAPO algorithm .
We test the DAG-based B2MAPO algorithm on the popular cooperative multi-agent benchmarks: StarCraftII Multi-agent Challenge (SMAC)\cite{Samvelyan_Rashid_Witt_Farquhar_Nardelli_Rudner_Hung_Torr_Foerster_Whiteson_2019} and Google Research Football (GRF)\cite{Kurach_Raichuk_Stańczyk_Zając_Bachem_Espeholt_Riquelme_Vincent_Michalski_Bousquet_et}. 
On all benchmarks, B2MAPO consistently outperforms strong baselines including MAPPO \cite{Yu_Velu_Vinitsky_Wang_Bayen_Wu_2021}, HAPPO \cite{Kuba_Chen_Wen_Wen_Sun_Wang_Yang_2021} and A2PO \cite{Wang_Tian_Wan_Wen_Wang_Zhang_2023} with a large margin in both performance and efficiency.

In summary, the main contributions of our work are as follows:

{

\setlength{\abovedisplayskip}{0pt}
\setlength{\belowdisplayskip}{0pt}
\begin{itemize} 
\setlength{\itemsep}{1pt}
\setlength{\parsep}{0pt}
\setlength{\parskip}{0pt}

\item To improve the efficiency of the model training, we propose a universal B2MAPO scheme. We prove that the guarantees of monotonic improvement of both joint and individual policies could be retained under the single rollout B2MAPO scheme with the off-policy correction method. We further prove that the joint monotonic bound is incrementally tightened when performing B2MAPO scheme. Our B2MAPO scheme is the first batch sequential update scheme.
\item  We design a universal modulized plug-and-play B2MAPO hierarchical framework, which adopts the CTDE principle. The B2MAPO framework supports the convenient integration of any MARL models with few or without modifications to fully exploit and merge their merits, including optimal performance and inference efficiency.
\item  Based on the B2MAPO framework, a carefully designed implementation of B2MAPO framework, called DAG-based B2MAPO algorithm, is devised. Experiments show that the B2MAPO algorithm outperforms state-of-the-art MARL methods and achieves both high efficiency and optimality.

\end{itemize}
}


\section{Problem Definition}
The decision-making problem of cooperative multi-agent can be formulated to a decentralized partially observable Markov decision process (Dec-POMDP)\cite{Bernstein_Givan_Immerman_Zilberstein_2002}, 
which is denoted by a tuple ${M\!=<\! I, S, \boldsymbol{A}, P, R, \Omega, \mathbb{T},\gamma \!>}$. 
$I \!=\! \{1, 2, \cdots,n\}$ is a finite set of agents. $S$ is the state space.
$\boldsymbol{A} \!\!=\!\! A^1 \! \times\! \cdots \!\times \! {A^n}$ is the joint action space, where $A^i$ is the action space of agent $i$. 
$P \! :\!  S \! \times \! \boldsymbol{A} \! \times \! S \! \rightarrow \! [0,1] $ is the transition function. 
All agents share the reward function $R: S \! \times \! \boldsymbol{A} \! \rightarrow \! \mathbb{R}$. 
$\Omega$ is the observation space.
$\mathbb{T} \equiv (\Omega \times \boldsymbol{A})^{*}$ is the space of action-observation trajectory. 
$\gamma \! \in \! [0,1)$ is a discounted factor.

At time step $t$, each agent $i$ obtains $o_t^i \!\in\! \Omega$ through the observation function $\Omega(s_t, i)$ and takes an action $a_t^i \!\!\in\!\! A^i$ based on its policy $\pi^i(\cdot|s_t)$, 
where the global state $s_t$ is approximated by the trajectory $\tau_t^i=\{o_1^i, a_1^i,\cdots, o_t^i, a_t^i\} \! \in \! \mathbb{T}$.
This forms the joint action $\boldsymbol{a}_t = \{a_t^1,\cdots,a_t^n\}$ and the joint policy $\boldsymbol{\pi} = (\pi^1,\cdots ,\pi^n)$. 
Then, the next state $s_{t+1}$ is obtained through the transition function $P(s_{t+1}|s_t,\boldsymbol{a}_t)$. 
The shared reward $r_t$ is obtained through the reward function $R(s_t,\boldsymbol{a}_t)$. 
The joint policy $\boldsymbol{\pi}$ induces a normalized discounted state visitation distribution $d^{\boldsymbol{\pi}}$, where $d^{\boldsymbol{\pi}}(s) = (1-\gamma)\sum_{t=0}^{\infty}\gamma^tPr(s_t= s|\boldsymbol{\pi})$ and $Pr(\cdot |\boldsymbol{\pi} ):S \mapsto [0,1]$ is the state probability function under $\boldsymbol{\pi}$.
The value function and advantage function are defined as $V^{\boldsymbol{\pi}}(s) \!= \! \mathbb{E}_{ (s_t,\boldsymbol{a}_t) \sim (P,\boldsymbol{\pi}) }[\sum_{t=0}^\infty\gamma^tR(s_t,\boldsymbol{a}_t)|s_0=s]$ and $A^{\boldsymbol{\pi}}(s, \boldsymbol{a}) = R(s, \boldsymbol{a})+\gamma\mathbb{E}_{s' \sim {P(\cdot|s,\boldsymbol{a})}}[V^{\boldsymbol{\pi}}(s')-V^{\boldsymbol{\pi}}(s)]$, respectively. 
The optimization objective of joint policy is to maximize the expected cumulative return, denoted as $\boldsymbol{\pi}^* = \arg\max_{\boldsymbol{\pi}}\mathcal{J}(\boldsymbol{\pi})=\arg\max_{\boldsymbol{\pi}}\mathbb{E}_{s_0 \sim d^{\boldsymbol{\pi}}, (s_t,\boldsymbol{a}_t ) \sim (P, \boldsymbol{\pi}})[\sum_{t=0}^\infty\gamma^tR(s_t, a_t)]$.

\section{Method}

To reduce the training and execution time(efficiency) while ensuring the optimal performance of B2MAPO, we first propose B2MAPO scheme in Section 3.1 and give its optimality theoretical guarantee. Based on B2MAPO scheme, a universal modulized plug-and-play B2MAPO hierarchical framework is designed in Section 3.2. Generating batches is a necessary step of B2MAPO scheme, while it is a NP-hard problem. Based on B2MAPO framework, we present a DAG-based B2MAPO approximate algorithm in Section 4 to solve the NP-hard problem efficiently.

\subsection{ Monotonic Improvement of B2MAPO scheme }

In order to improve the efficiency of the model training, we propose a \textbf{B}atch by \textbf{B}atch \textbf{M}ulti-\textbf{A}gent \textbf{P}olicy \textbf{O}ptimization (B2MAPO) scheme, which is a universal sequential batched policy updating approach.
Inspired by A2PO\cite{Wang_Tian_Wan_Wen_Wang_Zhang_2023}, two properties are proven to guarantee the optimal performance of B2MAPO scheme. (Proof details can be found in \cite{zhang2024b2mapo})

Given a set of agents $I$, in B2MAPO scheme, let $(B, \prec)$ denote the total order set of batches (sets of agents), called batch sequence, where $B\!=\! \{b_k | b_k \! \subseteq \! I \!\wedge \bigcup b_k\!=\!I \wedge \forall i \! \neq \! j, b_i \cap b_j \! = \! \varnothing \}$; 
$b_k$ represents the $k$-th batch which update its policies simultaneously;

$\prec$ is the binary relation on $B$, i.e. $b_i \! \prec \! b_j$ \emph{iff} $b_i$ is updated before $b_j$. 
Let $\pi^{b_k}$ and $\hat{\pi}^{b_k}$ denote the joint policy and the updated joint policy of $b_k$, respectively.  
Then, $\hat{\boldsymbol{{\pi}}}^{b_k} = \hat{\pi}^{b_1} \times \cdots \times \hat{\pi}^{b_{k}} \times {\pi}^{b_{k+1}} \times \cdots \times {\pi}^{b_{|B|}}$ indicates the joint policy of $B$ after $b_k$ is updated. 
In each round of policy updating, suppose ${\boldsymbol{\pi}}$ is the joint policy before $B$ is updated, while $\hat{\boldsymbol{\pi}}$ is the joint policy after $B$ is updated. 
Then, the batched policy updating process can be represented by:
{
\setlength{\parsep}{0pt}
\setlength{\parskip}{0pt}
\setlength{\abovedisplayskip}{3pt}
\setlength{\belowdisplayskip}{3pt}
\begin{eqnarray}
\boldsymbol{\pi}\! \xrightarrow[\text{Update }\pi^{b_1}\!]{\max\mathcal{L}_{\boldsymbol{\pi}}\!(\hat{\boldsymbol{\pi}}^{b_1}\!)}\!\hat{\boldsymbol{\pi}}^{b_1}\!\to\!\cdots\!\to\!\hat{\boldsymbol{\pi}}^{b_{{|B|}\!-\!1}}\! 
\xrightarrow[\text{Update }\pi^{b_{|B|}}]
{\max\mathcal{L}_{\hat{\boldsymbol{\pi}}^{b_{{|B|}\!-\!1}}}\!(\hat{\boldsymbol{\pi}}^{b_{|B|}}\!)}\!\hat{\boldsymbol{\pi}}^{b_{|B|}}\!=\!\hat{\boldsymbol{\pi}}\nonumber
\end{eqnarray}
}
where $\mathcal{L}_{\hat{\boldsymbol{\pi}}^{b_{k-1}}}(\hat{\boldsymbol{\pi}}^{b_k}) $ is the surrogate objective for batch $b_k$. 

While $\hat{\boldsymbol{\pi}}^{b_{k}}$ is updated from $\hat{\boldsymbol{\pi}}^{b_{k-1}}$, B2MAPO scheme uses the advantage function $A^{\hat{\boldsymbol{\pi}}^{b_{k-1}}}$ in the surrogate objective rather than $A^{\boldsymbol{\pi}}$.
Similar to the existing offline policy correction methods\cite{han2019dimension,han2023sample}, for each $b_k$, B2MAPO scheme employs importance sampling to correct the discrepancy between \(\hat{\boldsymbol{\pi}}^{b_{k-1}}\) and \(\boldsymbol{\pi}\). 
Thus, B2MAPO scheme still allows all batches to be updated through a single rollout.
Moreover, to maintain monotonic improvement and sample efficiency properties, the samples collected under \(\boldsymbol{\pi}\) are utilized to compute the estimated advantage function $A^{\boldsymbol{\pi},\hat{\boldsymbol{\pi}}^{b_{k-1}}}$ to approximate \(A^{\hat{\boldsymbol{\pi}}^{b_{k-1}}}\), employing truncated product importance weights to correct the probabilities:
{

\begin{eqnarray}
&A^{\hat{\boldsymbol{\pi}}^{b_{k-1}}}(s_{t},\boldsymbol{a}_{t}) \approx A^{\boldsymbol{\pi},\hat{\boldsymbol{\pi}}^{b_{k-1}}}(s_{t},\boldsymbol{a}_{t})=\delta_{t} + &\nonumber \\
&\sum_{n_{t}\geq1}\gamma^{n_t}(\prod_{j=1}^{n_t}\lambda\min{(1.0,\frac{\hat{\boldsymbol{\pi}}^{b_{k-1}}(\boldsymbol{a}_{t+j}|s_{t+j})}{\boldsymbol{\pi}(\boldsymbol{a}_{t+j}|s_{t+j})}}))\delta_{t+n_t}&
\end{eqnarray}
} 
where $\delta_{t} \!=\! r_t\!+\!\gamma V(s_{t+1})\!-\!V(s_t)$; $\lambda$ is a parameter which controls the bias and the variance; 
$\forall j \! \in \! \{1,\cdots,n_t\}$, $\min{(1.0,\frac{\hat{\boldsymbol{\pi}}^{b_{k-1}}(\boldsymbol{a}_{t+j}|s_{t+j})}{\boldsymbol{\pi}(\boldsymbol{a}_{t+j}|s_{t+j})}})$ is truncated importance sampling weights to approximate the probability of $s_{t+n_t}$ at time step $t + n_t$ under $\hat{\boldsymbol{\pi}}^{b_{k-1}}$.
Then, the surrogate objective of $b_k$ is:
{

\begin{eqnarray}
\!\mathcal{L}_{\hat{\boldsymbol{\pi}}^{b_{k\!-\!1}}}(\hat{\boldsymbol{\pi}}^{b_k})\!=\!\mathcal{J}(\hat{\boldsymbol{\pi}}^{b_{k\!-\!1}})\! 
+\!\frac{1}{1\!-\!\gamma}\mathbb{E}_{(s,\boldsymbol{a}\!)\!\sim\!(P,\hat{\boldsymbol{\pi}}^{b_k}\!)\!}A^{\boldsymbol{\pi}\!,\hat{\boldsymbol{\pi}}^{b_{k-1}}}(s,\boldsymbol{a})\!
\end{eqnarray}
}
The surrogate objective of $\boldsymbol{\pi}$ is:
{

\begin{eqnarray}
\mathcal{G}_{\boldsymbol{\pi}}(\hat{\boldsymbol{\pi}})=\mathcal{J}(\boldsymbol{\pi})+\frac{1}{1-\gamma}\sum_{k=1}^{|B|}\mathbb{E}_{(s,\boldsymbol{a})\sim(P,\hat{\boldsymbol{\pi}}^{b_k})}A^{\boldsymbol{\pi},\hat{\boldsymbol{\pi}}^{b_{k-1}}}(s,\boldsymbol{a})
\end{eqnarray}
}
\begin{theorem}
Given $(B,\prec)$, let $\alpha^{b_k}=D_{TV}^{\max}(\pi^{b_k}\|\hat{\pi}^{b_k})$, $\epsilon\!=\!\max_{b_k}\epsilon^{b_k}$, $\epsilon^{b_k}=\max_{s,\boldsymbol{a}}|A^{\hat{\boldsymbol{\pi}}^{b_{k-1}}}(s,\boldsymbol{a})|$, and $\xi^{b_k}=\max_{s,\boldsymbol{a}}|A^{\boldsymbol{\pi},\hat{\boldsymbol{\pi}}^{b_{k-1}}}(s,\boldsymbol{a})-A^{\hat{\boldsymbol{\pi}}^{b_{k-1}}}(s,\boldsymbol{a})|$, then: the monotonic bound of B2MAPO is incrementally tightened, i.e.
{
\begin{eqnarray}
|\nonumber\mathcal{J}(\hat{\boldsymbol{\pi}})-\mathcal{G}_{\boldsymbol{\pi}}(\hat{\boldsymbol{\pi}})|\leq\frac{\sum_{k\!=\!1}^{|B|}\xi^{{b_k}}}{1-\gamma}+ \enspace \enspace \enspace \enspace \enspace \enspace \enspace \enspace\\
4\epsilon\sum_{k\!=\!1}^{|B|}\alpha^{b_k}\!\left(\!\frac{1}{1\!-\!\gamma}\!-\!\frac{1}{1\!-\!\gamma\left(1\!-\!\sum_{j\in(B_k\!\cup\{\!b_k\!\})\!}\alpha^{j}\right)}\!\right)
\end{eqnarray}
}
\end{theorem}

\begin{theorem}

For any joint policy $\boldsymbol{\pi}_{A2PO}$, produced by A2PO, there exists the equivalent joint policy $\boldsymbol{\pi}_{B2MAPO}$ having the same expected discounted cumulative return, that can be acquired by B2MAPO scheme.
\end{theorem}

In order to prevent the non-stationary problem, all agents in the same batch must be independent. 
While the number of batches is minimized, the efficiency of policy optimization updating is maximized.
The minimum number of batches is the maximum number of agents that are interdependent.
However, to find the minimum number of batches is equivalent to the set coverage problem(SCP)\cite{caprara2000algorithms}, which is a NP-Hard problem.

Additionally, B2MAPO is a universal policy optimization scheme. 
Given ${M\!=<\! I, S, \boldsymbol{A}, P, R, \Omega, \mathbb{T},\gamma \!>}$ of $n$ agents, MAPPO can be regarded as B2MAPO with $|B| = 1$, while HAPPO and A2PO can be considered as  B2MAPO with $|B| = n$.
\\

\subsection{B2MAPO Framework}

Based on B2MAPO scheme, a universal modulized plug-and-play B2MAPO hierarchical framework, which adopts the CTDE principle, supports convenient integration of any MARL models to fully exploit and merge their merits, including optimal performance and inference efficiency.
The structure of B2MAPO framework is illustrated in Figure 2. 
It is a two-layer hierarchical structure, consisting of a batch sequence generation layer(upper layer) and a batch by batch policy optimization layer(lower layer).
The upper layer contains two modules, batch partitioning and batch ordering, which partition agents to batches and generate batch updating order.  
The lower-layer includes two joint policy networks, a joint policy $\boldsymbol{\pi}$ trained by B2MAPO scheme and a derived joint policy $\boldsymbol{\pi}_{ind}$ according to CTDE-based MARL algorithms.

\begin{figure}[ht]
\begin{center}
\centerline{\includegraphics[width=0.8\columnwidth]{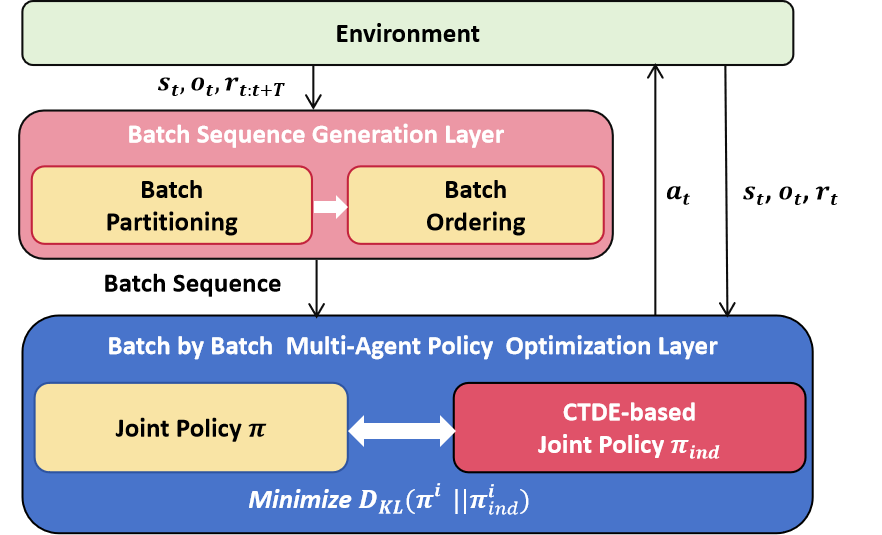}}
\caption{The structure of B2MAPO framework.}
\end{center}

\label{framework}
\end{figure}

During centralized training, at the time step $t$, the upper-layer periodically outputs batch sequence based on $s_t$ for future $T$ time steps. 
Batch by batch policy optimization layer utilizes the batch sequence to train $\boldsymbol{\pi}$ and $\boldsymbol{\pi}_{ind}$ in parallel.
$\boldsymbol{\pi}$ and $\boldsymbol{\pi}_{ind}$ receive $s_t$ and joint observation $\boldsymbol{o}_t$ to generate joint actions $\boldsymbol{a}_t$. 
$\boldsymbol{\pi}$ performs MARL training based on the reward $r_t$.
During the execution phase, only $\boldsymbol{\pi}_{ind}$ is employed for decision-making. 
At each time step $t$, $\boldsymbol{\pi}_{ind}$ outputs joint actions $\boldsymbol{a}_t$ based on joint observations $\boldsymbol{o}_t$.

\subsubsection{Batch Sequence Generation Layer}

Given a time period $T$, batch sequence generation layer periodically produces optimal batch sequence $(B,\prec)^{OPT}$.
Based on $s_t$ and $r_t$, batch partitioning module tries to divide all agents into minimum number of batches, while batch ordering module produces the most appropriate updating order for partitioned batches.

If agents in the same batch have dependencies, it will lead to a non-stationary problem, so the degree of dependency is learned for batch partition.
In practice, the policy dependency of agents is very hard to be obtained. 
Nevertheless, the degree of the policy interdependence between agents can be revealed through deliberately designed networks like attention mechanisms\cite{vaswani2017attention}. 
Then, several applicable methods can be employed to generate batches.
For fewer agents, batches can be partitioned by brute force search\cite{schaeffer1993re}; 
Otherwise, for a greater number of $n$ agents, the batch partition can be solved by any approximate algorithm of SCP\cite{hochba1997approximation} in polynomial time complexity $\Theta(logn)$.
Additionally, if the number of batches is determined, agents can be partitioned into batches randomly.
The influence on the quality of batch partition can be analyzed by estimating the probability that any related agents are grouped into the same batch.

Theoretically, for correctly partitioned batches, their updating order does not affect the improvement guarantee of joint policy.
However, due to errors in estimating the advantage function and surrogate objective, the joint policy may still fall into a sub-optimum, and the optimality cannot be guaranteed in practice. 
A2PO\cite{Wang_Tian_Wan_Wen_Wang_Zhang_2023} alleviates this problem by sequential updates in descending order of the estimated advantage function.
As study in \cite{Wang_Tian_Wan_Wen_Wang_Zhang_2023}, 
while training a decision model, a better sequential policy update order can produce a joint policy with greater performance. 

Additionally, the dependent relationships among agents can be modeled as a directed or undirected agents' interdependent graph, in which vertex and edges represent agents and their dependency, respectively.
Although, generally, the dependency among agents can not be explicitly determined, 
the SARL algorithms combined with attention mechanism\cite{vaswani2017attention} can be utilized to discover the policy dependency more precisely.
Then, a spanning subgraph of the agents' interdependent graph can be generated by many algorithms, such as DAG-notear\cite{Zheng_Aragam_Ravikumar_Xing_2018}, DAG-GNN\cite{Yu_Chen_Gao_Yu_2019}, and DAG-nocurl\cite{Yu_Gao_Yin_Ji_2021}.
In section 4.1, we propose a hierarchical DAG generation method to produce this spanning subgraph.
Finally, the batch sequence of agents can be computed based on the spanning subgraph by topological sorting.

\subsubsection{Batch by Batch Policy Optimization Layer}

To speed up the convergence rate and improve the performance of decision model,
additional information, e.g. action sequences of preceding agents, is taken as a part of joint policy input, such as RPISA.
However, those approaches do not satisfy CTDE principle. 
Moreover, the complexity of those models, such as RPISA, HAPPO and A2PO, might increase inference costs. 
On the other hand, the CTDE-based models with lightweight networks, e.g. MAPPO, have higher inference efficiency, but cannot ensure optimal performance.

$\boldsymbol{\pi}_{ind}$, which is composed of independent policies of all agents, inferences efficiently with lightweight networks and satisfies CTDE principle. $\boldsymbol{\pi}$ supports B2MAPO scheme and guarantees optimal performance.
The CTDE-based derived joint policy $\boldsymbol{\pi}_{ind}$ is trained with different experiences gathered by joint policy $\boldsymbol{\pi}$ that is produced by B2MAPO scheme.
Meanwhile, according to theorem \ref{theorem 3}, $\boldsymbol{\pi}_{ind}$ is derived from $\boldsymbol{\pi}$ by periodically minimize the KL divergence between $\boldsymbol{\pi}_{ind}$ and $\boldsymbol{\pi}$. (Proofs can be found in \cite{zhang2024b2mapo})

\begin{theorem}
\label{theorem 3}
For any joint policy $\boldsymbol{\pi}$ produced by B2MAPO scheme, there exists a joint policy $\boldsymbol{\pi}_{ind}$ acquired by CTDE-based MARL algorithms that does not involve dependency among agents, such that $V_{\boldsymbol{\pi}}(s) = V_{\boldsymbol{\pi}_{ind}}(s)$ for any state $s \in S$.
\end{theorem}

B2MAPO framework provides a universal modulized plug-and-play architecture, which can conveniently integrate third-party models with few or even without modification to exploit their merits.
"CTDE-based joint policy" module in this layer can be implemented by various MARL methods, such as Q-learning-based methods\cite{Rashid_Samvelyan_Witt_Farquhar_Foerster_Whiteson_2018, Son_Kim_Kang_Hostallero_Ye_2019, Wang_Ren_Liu_Yu_Zhang_2020}, policy-based methods\cite{Yu_Velu_Vinitsky_Wang_Bayen_Wu_2021}, and actor-critic methods\cite{Yuan_Wang_Wang_Zhang_Chen_Guan_Zhang_Zhang_Yu}. 

{

\section{DAG-based B2MAPO Algorithm}
}
{
\begin{figure*}[htbp]
\centering
\includegraphics[width=0.85\linewidth]{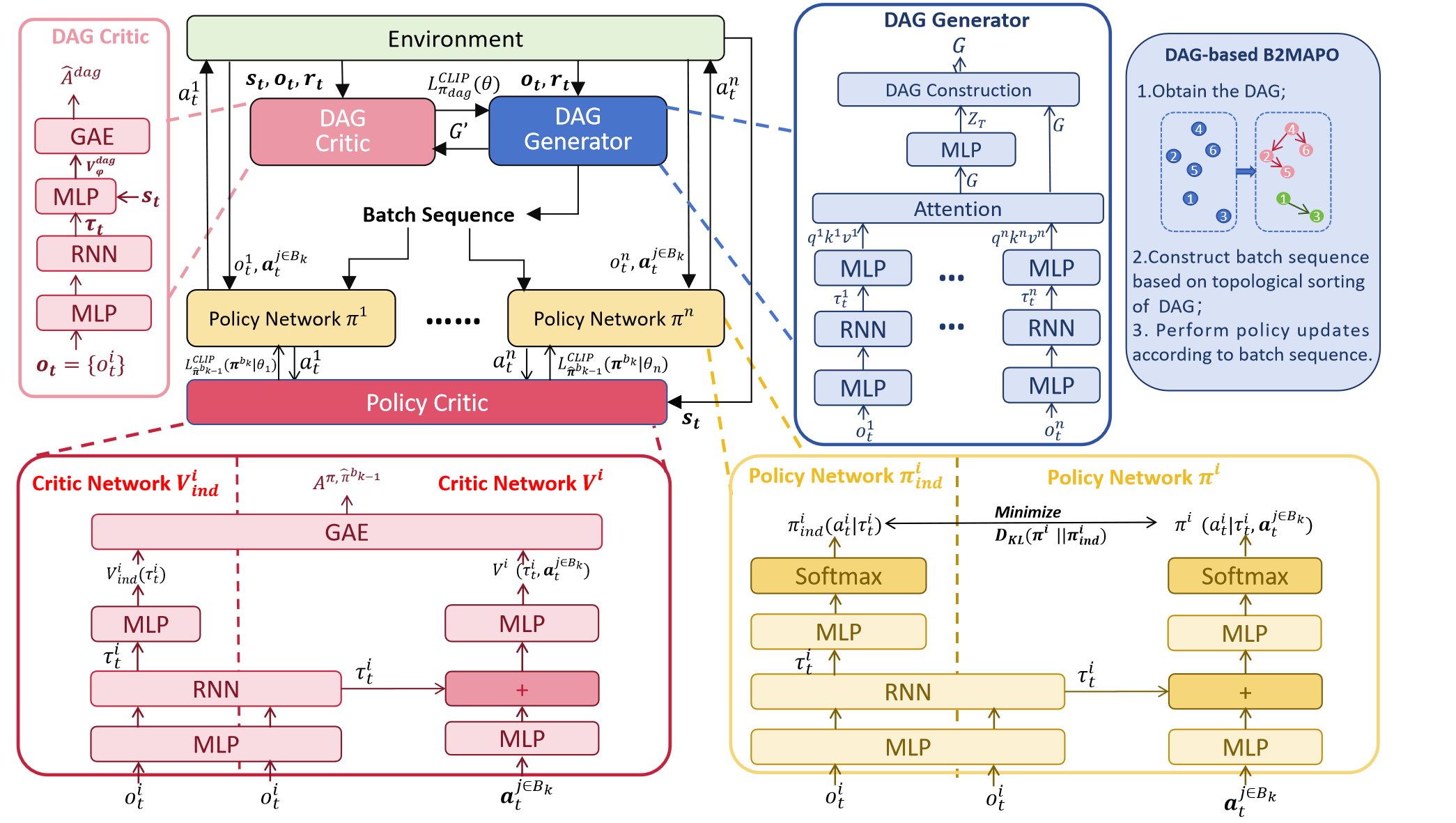}

\caption{The network architecture of the DAG-based B2MAPO algorithm.}

\label{architecture}
\end{figure*}
}

DAG-based B2MAPO algorithm is a carefully designed implementation of the universal B2MAPO framework.
The upper layer contains a DAG critic and a DAG generator trained by PPO algorithm\cite{schulman2017proximal}.
The DAG generator constructs DAGs of agents to produce optimal batch sequence.
In the lower layer, we design a policy network for an agent, which can be shared among all agents.
Joint policy $\boldsymbol{\pi}$ is trained based B2MAPO scheme with optimal batch sequence, while the derived joint policy $\boldsymbol{\pi}_{ind}$ is updated by MAPPO.
Figure \ref{architecture} presents the network architecture of the DAG-based B2MAPO algorithm.
\\
{
\label{4.1}

\subsection{ Hierarchical DAG Generation}

}

The DAG generator utilizes attention mechanism build agent interdependent graph by encoding historical trajectories every $T$ time steps.
DAG construction module periodically collects the graph to produce its spinning subgraph (DAG of agents).
The optimal batch sequence is generated based on the DAG by topological sorting.

During centralized training, at time step $t$, for each agent $i$, DAG generator encodes the observation $o_t^i \! \in \! \boldsymbol{o}$ into the historical trajectory $\tau_t^i \! \in \! \boldsymbol{\tau}_t$ by multilayer perceptron(MLP) and recurrent neural network (RNN) module. 
While the historical trajectory $ \tau^{i}_{t}$ of each agent $i$ is encoded, the query vectors $q^i = \textbf{W}_q\tau^{i}_{t}$ , the key vectors $k^i=\textbf{W}_k\tau^{i}_{t}$ and the value vectors $v^i=\textbf{W}_v\tau^{i}_{t}$ are calculated as well. 
Then, the graph $G_T = [G^1_T, G^2_T, \ldots, G^n_T]^\top$ represented the interdependence between agents is derived, 
where $G^i_T$ signifies the degree to which agent $i$ depend on other agents.
{
\setlength{\abovedisplayskip}{3pt}
\setlength{\belowdisplayskip}{3pt}
\begin{eqnarray}
    G^i_T= [g_T^{ij}] =\text{softmax}\left[\frac{q_i^\top k_1}{\sqrt{d_k}},\frac{q_i^\top k_2}{\sqrt{d_k}},\ldots,\frac{q_i^\top k_n}{\sqrt{d_k}}\right]
\end{eqnarray}
}
\\
where $g_T^{ii}=0$. 
In addition, if $g_T^{ij}<\delta $, $g_T^{ij}=0$,  where $ \delta >0$ is a hyper-parameter indicated the minimum degree of interdependence.

\begin{figure*}[htbp]
    \centering
    \begin{minipage}[t]{0.67\linewidth}
    \centering
    \includegraphics[width=0.32\linewidth]{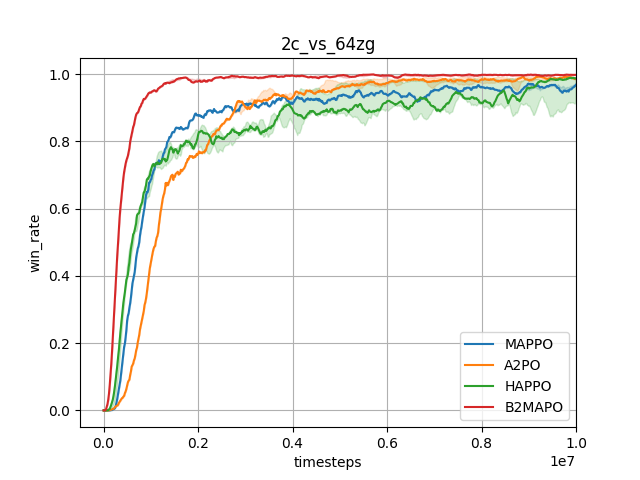}
    \includegraphics[width=0.33\linewidth]{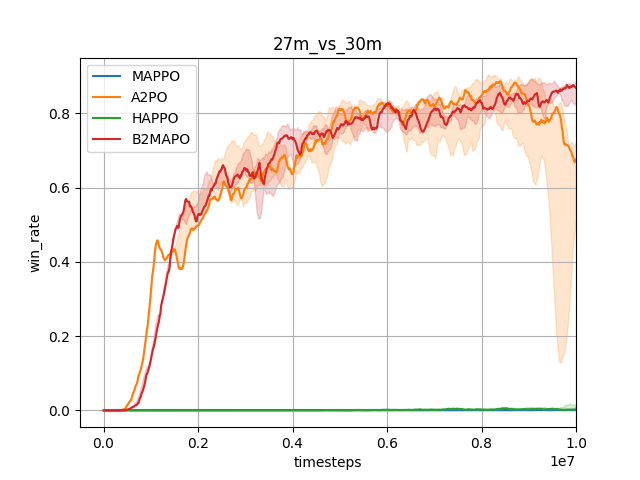}
    \includegraphics[width=0.33\linewidth]{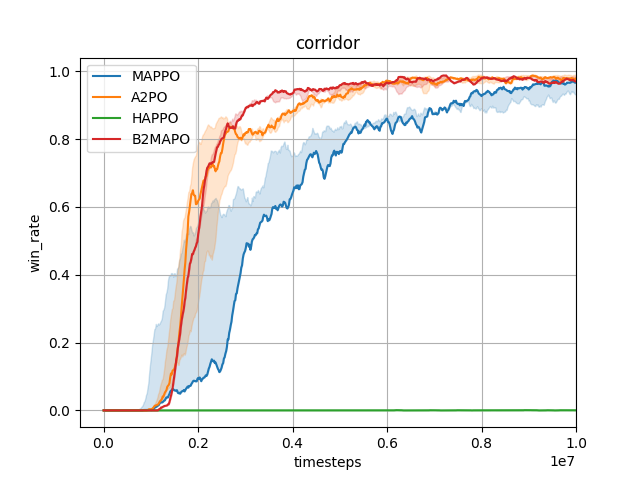}\\
    \includegraphics[width=0.32\linewidth]{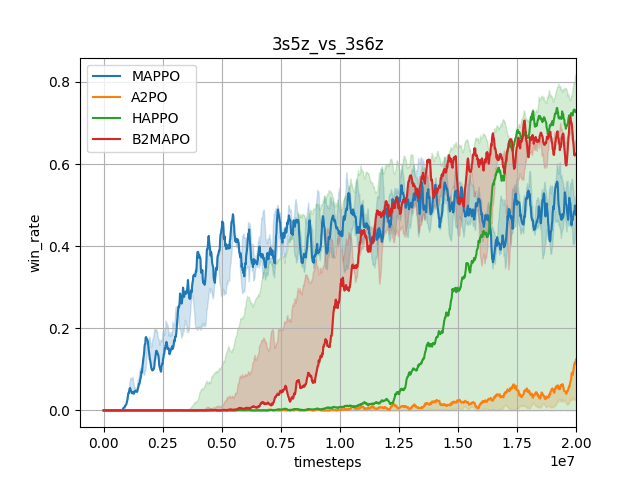}
    \includegraphics[width=0.33\linewidth]{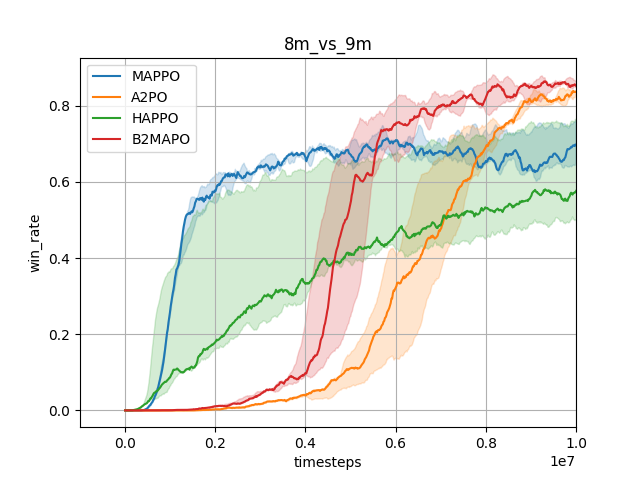}
    \includegraphics[width=0.33\linewidth]{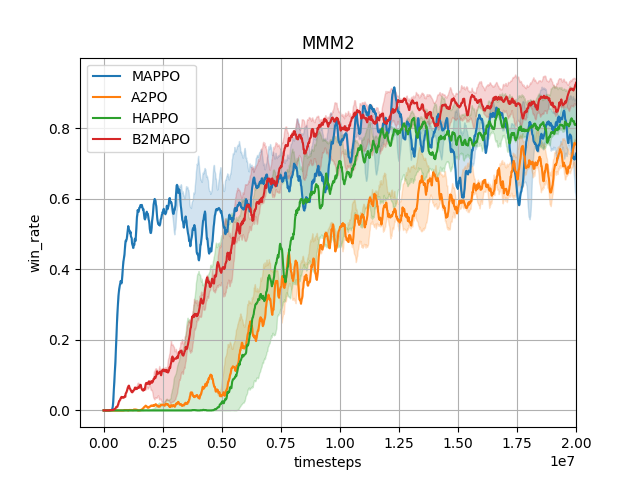}
    \caption{The comparison of performance between B2MAPO and other baselines in SMAC.}
    \label{fig:ex1}
    \end{minipage}
    \centering
    \begin{minipage}[t]{0.32\linewidth}
    \centering
\includegraphics[width=0.64\linewidth]{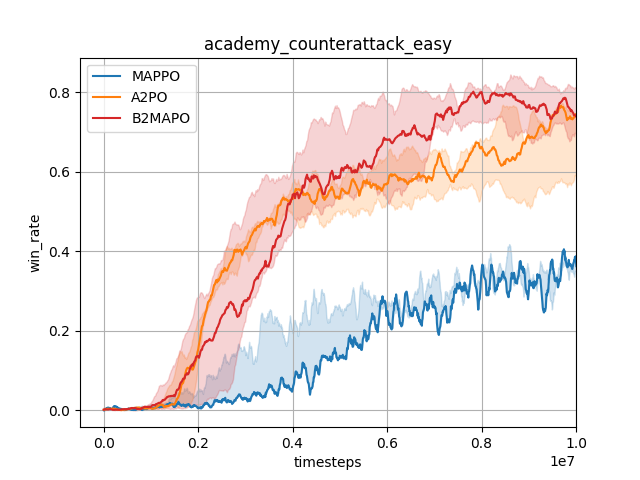}
\includegraphics[width=0.64\linewidth]{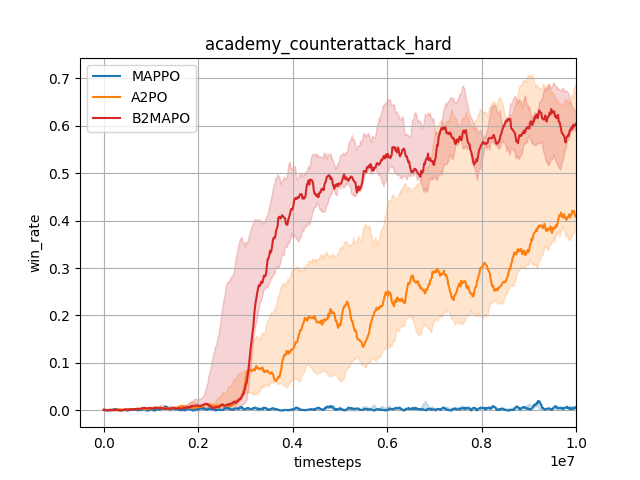}
\caption{The comparison of performance between B2MAPO and other baselines in GRF.}
\label{ex:grf}
\end{minipage}

\end{figure*}
The graph $G_T$ is embedded by an MLP to obtain an abstract representation $Z_T = MLP(v^i{G^i_T})$, which is utilized to generate and initialize the parameters of DAG construction module.
The DAG construction module is implemented based on \cite{Cordeiro_Mounié_Perarnau_Trystram_Vincent_Wagner_2010}
DAG construction module outputs the spinning subgraph $G$ based on $G_T$.
The batch sequence $(B,\prec)$ for the future $T$ steps is obtained by topological sorting of the spinning subgraph DAG $G$.

DAG Critic is employed to estimate the advantage function $\hat{A}^{dag}$ with $(B,\prec)$ from time step $t$ to $t+T$; 
Similar to DAG Generator, at time step $t$, DAG Critic encodes joint observation $\boldsymbol{o}_t$ into joint historical trajectory $\boldsymbol{\tau}_t$, through MLP and RNN module.
DAG Critic generates the estimated state value $V_\phi^{dag}(s_t)$ based on the embedding of the global state $s_t$ and joint historical trajectory $\boldsymbol{\tau}_t$.
Finally, DAG Critic, combined with Generalized Advantage Estimation (GAE)\cite{schulman2015high} module for more precise estimation, approximates the advantage function $\hat{A}^{dag}$ for the future $T$ steps:

{
\setlength{\abovedisplayskip}{0pt}
\setlength{\belowdisplayskip}{0pt}
\begin{eqnarray}
    \hat{A}^{dag} = \sum_{l=0}^h(\gamma)^{l}\delta_{t+l}^{dag}
\end{eqnarray}
}
{
\setlength{\abovedisplayskip}{2pt}
\setlength{\belowdisplayskip}{2pt}
\begin{eqnarray}
    \delta^{dag}_{t} = r_{t:t+T} + \gamma V_{\phi}^{dag}(s_{t+T}) - V_{\phi}^{dag}(s_{t})
\end{eqnarray}
}

The upper-layer network is trained based on maximizing the discounted cumulative reward $r_{t:t+T} = \sum_{l=0}^{T}(\gamma)^{l}R(s_{t+l},\boldsymbol{a}_{t+l})$ over a fixed period $T$. 
The loss function of DAG Generator is :
{
\setlength{\abovedisplayskip}{2pt}
\setlength{\belowdisplayskip}{2pt}
\begin{eqnarray}
    &L(\theta) = \mathbb{E}_t \left[ \min\left( Sr(\theta) \hat{A}^{dag}, \text{clip}(Sr(\theta), 1-\epsilon, 1+\epsilon)\hat{A}^{dag} \right) \right]& \nonumber \\
&- c_1 \mathbb{E}_t \left[ D_{KL} \left( \pi^{dag}_{\theta_{\text{old}}}(\cdot | s_t) \| \pi^{dag}_{\theta}(\cdot | s_t) \right) \right] \nonumber& 
\end{eqnarray}
}
where $\pi^{dag}_{\theta}$ is the current DAG generator, $\pi^{dag}_{\theta_{old}}$ is the target DAG generator,  $Sr(\theta) = \frac{\pi^{dag}_{\theta}(a_t | s_t)}{\pi^{dag}_{\theta_{old}}(a_t | s_t)}$ is the important sample rate between $\pi^{dag}_{\theta}$ and $\pi^{dag}_{\theta_{old}}$.

DAG critic is trained to minimize the loss function:
{

\begin{eqnarray}
    &L(\phi) = \mathbb{E}_t \left[ \left( r_{t:t+T} - V_{\phi}^{dag}(s_{t}) \right)^2 \right]&
\end{eqnarray}
}
where $V_{\phi}^{dag}(s_{t})$ is the DAG critic to estimate state value for the next $T$ steps, which is parameterized by $\phi$.\\
\subsection{Joint Policies Optimization}
As shown in Figure 3, the right side of the dotted line illustrates the critic $V^i$ and policy $\pi^i$ networks of agent $i$ corresponding to joint policy $\boldsymbol{\pi}$. 
The left side of the dotted line illustrates the critic $V^i_{ind}$ and policy $\pi^i_{ind}$ networks of agent $i$ corresponding to joint policy $\boldsymbol{\pi}_{ind}$.
$V^i_{ind}$ and $\pi^i_{ind}$ is implemented according to MAPPO\cite{Yu_Velu_Vinitsky_Wang_Bayen_Wu_2021}.

Different from the implementation of MAPPO, $V^i$ and $\pi^i$ of B2MAPO are additionally able to access the action sequence $\boldsymbol{a}_t^{j \in B_k}$ of preceding batch agents in addition to its own observations $o_t^i$. 
During centralized training, $V^i$ and $\pi^i$ are sequentially updated with batch sequence $(B,\prec)$.
Similar to the upper layer network, at timestep $t$, $V^i$ and $\pi^i$ obtain the historical trajectory $ \tau^{i}_{t}$ by embedding the $ o^{i}_{t}$ through MLP and RNN models.
$V^i$ encodes the behavior sequences $\boldsymbol{a}_t^{j \in B_k}$ of preceding batch agents and historical trajectory $ \tau^{i}_{t}$ to obtain the value  estimation $V^i(s_t)$ of the current state $s_t$. 
These value estimations assist in appropriating the advantage function $A^{\boldsymbol{\pi},\hat{\boldsymbol{\pi}}^{b_{k-1}}}(s,\boldsymbol{a})$ through the GAE module, as described in Section 3.1.
Similar to $V^i$, $\pi^i$ also utilizes the encoded features of the action sequences $\boldsymbol{a}_t^{j \in B_k}$ of preceding batch agents and historical observation trajectory $ \tau^{i}_{t}$, which are then passed through a softmax network to obtain the action probability distribution $\pi^i(\cdot|\tau^{i}_{t},\boldsymbol{a}_t^{j \in B_k})$.

To ensure the monotonic improvement of the joint policy $\boldsymbol{\pi}$, when updating $b_k$, we control the monotonic bound by minimizing the total variation distance $\alpha^{b_k}\sum_{{b_j}\in(B_k\cup\{b_k\})}\alpha^{b_j}$. 
Consequently, the surrogate objective of batch $b_k$ is:
{
\setlength{\abovedisplayskip}{2pt}
\setlength{\belowdisplayskip}{2pt}
\begin{eqnarray}
    &\mathcal{L}_{\hat{}{\boldsymbol{\pi}}^{b_{k-1}}}(\hat{\boldsymbol{\pi}}^{b_k}) = \mathcal{J}(\hat{\boldsymbol{\pi}}^{b_{k-1}})& \nonumber\\ &+\frac1{1-\gamma}\mathbb{E}_{(s_t,\boldsymbol{a}_t)\sim(d^{\boldsymbol{\pi}},\boldsymbol{\pi})}[\frac{\hat{\pi}^{b_k}\prod_{b_j\in B_k}\hat{\pi}^b_j}{\pi^{b_k}\prod_{b_j\in B_k}\pi^b_j}A^{\boldsymbol{\pi},\hat{\boldsymbol{\pi}}^{b_{k-1}}}(s_t,\boldsymbol{a}_t)]&
\end{eqnarray}
}
where $\mathcal{J}(\hat{\boldsymbol{\pi}}^{b_{k-1}})$ has no dependence to the policies of batch $b_k$. 
To improve the performance effectively, we consider applying the clipping mechanism to the joint policy ratio $\frac{\hat{\pi}^{b_k}\prod_{b_j\in B_k}\hat{\pi}^{b_j}}{\pi^{b_k}\prod_{b_j\in B_k}\pi^{b_j}}$.
The clipping mechanism is applied to joint policy ratio of $B_k$ and joint policy ratio of $b_k$ to reduce the instability of estimating the surrogate objective of $b_k$.
Finally, the surrogate objective for updating $b_k$ becomes: 
{
\setlength{\abovedisplayskip}{2pt}
\setlength{\belowdisplayskip}{2pt}
\begin{eqnarray}
    &\tilde{\mathcal{L}}_{\hat{\boldsymbol{\pi}}^{b_{k-1}}}(\hat{\boldsymbol{\pi}}^{b_k})=\mathbb{E}_{(s_t,\boldsymbol{a}_t)\sim(d^{\boldsymbol{\pi}},\boldsymbol{\pi})}
    \big[\min(l(s_t,\boldsymbol{a}_t)A^{\boldsymbol{\pi}, \hat{\boldsymbol{\pi}}^{b_{k-1}}},& \nonumber\\
&\operatorname{clip}(l(s_t,\boldsymbol{a}_t),1\pm\epsilon^{b_k})A^{\boldsymbol{\pi},\hat{\boldsymbol{\pi}}^{b_{k-1}}}) \big] &
\
\end{eqnarray}
}
{
\setlength{\abovedisplayskip}{3pt}
\setlength{\belowdisplayskip}{3pt}
\begin{eqnarray}
l(s_t,\boldsymbol{a}_t)=\frac{\hat{\pi}^{b_k}(\boldsymbol{a}_t^{b_k}|s_t)}{\pi^{b_k}(\boldsymbol{a}^{b_k}_t|s_t)}g(s_t,\boldsymbol{a}_t)
\end{eqnarray}
}
{
\setlength{\abovedisplayskip}{3pt}
\setlength{\belowdisplayskip}{3pt}
\begin{eqnarray}
g(s_t,\boldsymbol{a}_t)=\operatorname{clip}(\frac{\prod_{{b_j}\in B_k}\hat{\pi}^{b_j}(\boldsymbol{a}^{b_j}|s_t)}{\prod_{b_j\in B_k}\pi^{b_j}(\boldsymbol{a}^{b_j}|s_t)},1\pm\frac{\epsilon^{b_k}}2)
\end{eqnarray}
}
where $\epsilon^{b_k}$ is the clipping parameter.

To satisfy the CTDE principle, we derive a joint policy $\boldsymbol{\pi}_{ind}$ through diverse experiences explored by $\boldsymbol{\pi}$ and periodically minimize the KL divergence between $\boldsymbol{\pi}_{ind}$ and $\boldsymbol{\pi}$. 
Ultimately, the agent $i$ for execution only with the derived policy ${\pi}_{ind}^i$ has higher efficiency and optimal performance. 
The additional loss function of training the joint policy ${\pi}_{ind}$ is as follows:
{
\setlength{\abovedisplayskip}{2pt}
\setlength{\belowdisplayskip}{2pt}
\begin{eqnarray}
    L_{\boldsymbol{\pi}^{ind}} = \mathbb{E}_t\left[ D_{KL}( \pi^i \| \pi_{ind}^i)\right] , \forall i \in I
\end{eqnarray}
}
\begin{table*}[ht]
\caption{The comparison of average training time (seconds) between B2MAPO and other baselines.}

\centering
\begin{tabular*}{0.88\linewidth}{llll@{\hspace{8mm}}llll} 
\toprule
Algorithm & MMM2 & corridor&3s5z\_vs\_3s6z& 6h\_vs\_8z& 27m\_vs\_30m&2c\_vs\_64zg  &Average time \\ 
&(10 Agents) &(6 Agents)&(8 Agents)&(6 Agents)&(27 Agents)&(2 Agents) &(Compare To A2PO) \\
\midrule
A2PO &  \textbf{\textcolor{red}{0.02941030}} &   \textbf{\textcolor{red}{0.02880984}} &   \textbf{\textcolor{red}{0.04470281}} &   \textbf{\textcolor{red}{0.02794014}} &   \textbf{\textcolor{red}{0.098170261}}&   \textbf{\textcolor{red}{0.01103425}} & \textbf{\textcolor{red}{0.040011267}}\\
MAPPO &  0.00209796  & 0.00512125 &  0.00524281&  0.00426739&0.00479105&  0.00400626 & 0.004254453\(\downarrow 89.4\%\)\\
B2MAPO &  0.00666901& 0.01440159 & 0.02048663& 0.00951390& 0.03716109& 0.00692575 &0.015859662\(\downarrow 60.4\%\)\\
\bottomrule

\label{table1}
\end{tabular*}

\end{table*}

\begin{table*}[ht]

\caption{The comparison of average execution time (seconds) between B2MAPO and other baselines.}

\centering
\begin{tabular*}{0.88\linewidth}{llll@{\hspace{8mm}}llll} 
\toprule
Algorithm & MMM2 & corridor&3s5z\_vs\_3s6z& 6h\_vs\_8z& 27m\_vs\_30m&2c\_vs\_64zg  &Average time \\
&(10 Agents) &(6 Agents)&(8 Agents)&(6 Agents)&(27 Agents)&(2 Agents) &(Compare To A2PO) \\
\midrule
A2PO &  \textbf{\textcolor{red}{0.00783219}} &  \textbf{\textcolor{red}{0.00526356}} & \textbf{\textcolor{red}{0.00706817  }}&  \textbf{\textcolor{red}{0.00583951}}  &  \textbf{\textcolor{red}{0.01898989}} & \textbf{\textcolor{red}{0.00368452}}  & \textbf{\textcolor{red}{0.008112973}}
\\

MAPPO &  0.00179175 & 0.00165824 &  0.00154937&  0.001744142&0.00162831&  0.00110432 &0.001579355\(\downarrow 80.5\%\)
\\

B2MAPO &  0.00189891& 0.00160286 & 0.00188361& 0.00181356& 0.00188927& 0.00175449& 0.001807117\(\downarrow 78.7\%\)
\\
\bottomrule

\label{table2}
\end{tabular*}
\end{table*}

\section{Experiments}

This section empirically evaluates the DAG-based B2MAPO( abbreviated as B2MAPO) on the widely adopted collaborative multi-agent benchmark, the StarCraftII Multi-agent Challenge (SMAC) \cite{Samvelyan_Rashid_Witt_Farquhar_Nardelli_Rudner_Hung_Torr_Foerster_Whiteson_2019} and Google Football Research (GRF)\cite{Kurach_Raichuk_Stańczyk_Zając_Bachem_Espeholt_Riquelme_Vincent_Michalski_Bousquet_et}. 
We compare B2MAPO with advanced MARL methods: MAPPO \cite{Yu_Velu_Vinitsky_Wang_Bayen_Wu_2021}, HAPPO \cite{Kuba_Chen_Wen_Wen_Sun_Wang_Yang_2021}, and A2PO \cite{Wang_Tian_Wan_Wen_Wang_Zhang_2023}. All Results are average win rates of 5 random seeds.
For consistency, we implement all the algorithms with parameter sharing in SMAC and parameter-independent in GRF, according to the homogeneity and heterogeneity of agents. Experimental results demonstrate that:
1)B2MAPO outperforms state-of-the-art MARL methods in performance and efficiency. 
2)B2MAPO can stimulate coordinated behaviors and has significant advantages in complex cooperative tasks.

\subsection{ Performance Analysis}
\textbf{StarCraftII Multi-agent Challenge (SMAC).} We compared B2MAPO with advanced MARL baselines on the super hard maps MMM2, corridor, 3s5z\_vs\_3s6z, 27m\_vs\_30m, the hard map 2c\_vs\_64zg and 8m\_vs\_9m. 
All experiments were conducted using 5 different random seeds for evaluation purposes. 
Results in Figure \ref{fig:ex1} show that B2MAPO outperforms the baselines with better stability and achieves higher sample efficiency in all maps.

MAPPO and HAPPO fall into the dilemma of sub-optimality in some complex scenarios, such as 8m\_vs\_9m and 3s5z\_vs\_3s6z map. 
In the hard map 8m\_vs\_9m, although MAPPO and HAPPO converge faster in the early stage, they ultimately perform poorly in global target optimization. 
Since MAPPO and HAPPO lack a theoretical guarantee for the policy improvement of each agent, which causes the MAPPO and HAPPO policies are overfitted in the early stage which makes it difficult to learn the optimal policy.
According to the experimental results, the HAPPO algorithm is more unstable in complex scenarios, which is reflected in the high variance of HAPPO strategy training, especially in 3s5z\_vs\_3s6z, 8m\_vs\_9m, and MMM2 maps.
In challenging scenarios, such as the corridor, and 27m\_vs\_30m map, the HAPPO method performs poorly, exhibiting a zero win rate.
Under the guidance of joint policy optimization, HAPPO can occasionally avoid overfitting, and in 3s5z\_vs\_3s6z, its performance slightly surpasses B2MAPO.
Particularly, the performance of our B2MAPO algorithm is comparable to A2PO in all scenarios and even outperforms the A2PO in some complex scenarios.
In most cases, B2MAPO can achieve faster convergence, reflecting higher sample efficiency. 
However, in map 8m\_vs\_9m, 3s5z\_vs\_3s6z, and MMM2, B2MAPO converges slowly in the early stage. 
This may be because hierarchical reinforcement learning requires more time for the upper layer network to learn effectively, but in the end, the overall performance of B2MAPO is the best among all baselines.
Overall, our B2MAPO method outperforms the baseline methods regarding win rates across all scenarios, partly due to the utilization of dependencies among agents to generate batch sequences effectively for policy optimization. 

\textbf{Google Football Research(GRF)}
In this section, we test B2MAPO on two challenging GRF scenarios:
academy\_counterattack\_easy and academy\_counterattack\_hard scenarios, which pose challenges for agents in discovering complex coordination behaviors. 
We compare performance against baselines in Figure \ref{ex:grf}. 
MAPPO struggles to learn sophisticated policies in all scenarios. 
Due to heuristic order sequence generation, as the difficulty of the experiment increases, the performance of the A2PO algorithm deteriorates, and the instability increases, which is reflected in the high variance and low winning rate in the academy\_counterattack\_hard scenario.
Our B2MAPO outperforms all baselines in all scenarios with a stable training process, demonstrating superior performance in complex cooperative tasks.
Especially in the difficult academy\_counterattack\_hard scenario, B2MAPO has a higher and more stable performance, far better than A2PO, mainly reflected in small variance and high winning rate.

\subsection{Efficiency Analysis}
Table \ref{table1} records the average training time for all agents of each round in different SMAC maps. 
We found that B2MAPO ensures optimal performance while its training time is far less than A2PO (batch size = 1) and slightly higher than MAPPO (batch size = n). 
Specifically, in most scenarios such as MMM2, the training time of B2MAPO is significantly less than A2PO and close to MAPPO. 
Overall, B2MAPO reduced the average training time of A2PO by 60.4\%, and MAPPO reduced it by 89.4\%. 
This can be attributed to B2MAPO scheme, which effectively reduces training time. 
However, in complex scenarios like corridor, the training time of B2MAPO is close to A2PO, possibly due to the complexity of the environment. 
As shown in Table \ref{table1}, the training time of B2MAPO is directly proportional to the number of batches but not linearly so. 
This is due to the additional computations required by the B2MAPO algorithm, such as policy correction, clipping mechanism, and other calculations not required in regular MAPPO. 
As B2MAPO can adjust the batch sequence for policy optimization in real-time dynamically, it can maintain performance superiority while showing relatively high training efficiency.

Further, we conducted a comparison of the average execution time. 
The results are shown in table \ref{table2}. 
Both B2MAPO and MAPPO reduce the execution time of A2PO by about 80\%. 
Since B2MAPO uses a joint policy implemented by MAPPO for decision-making, its execution time is comparable to MAPPO. 
On the other hand, A2PO, due to the complexity of its network design, results in significantly higher execution time even if all agents make decisions simultaneously, far exceeding those of B2MAPO and MAPPO algorithms.

Although MAPPO has the shortest training and execution time, its performance is not guaranteed to be optimal.
On the other hand, A2PO can guarantee policy optimality, but it has poor training and execution efficiency.
Experimental results show that B2MAPO can achieve high training and execution efficiency while ensuring optimal performance.

\section{Related Works}

Hierarchy\cite{Dayan_Hinton_1992, Sutton_Precup_Singh_1999} has recently started to fulfill its potential for sample-efficient learning in challenging, long-horizon tasks. 
Recently, the concept of Hierarchical reinforcement learning (HRL)\cite{Dwiel_Candadai_Phielipp_Bansal_2019, Vezhnevets_Osindero_Schaul_Heess_Jaderberg_Silver_Kavukcuoglu_2017, Nachum_Gu_Lee_Levine_2018, Levy_Konidaris_Platt_Saenko_2017} revolves around decomposing a complex problem into a hierarchy of more manageable subtasks.
Initial studies in hierarchical MARL utilizes hand-crafted subtasks or sub-goals to facilitate cooperation among agents at different levels of abstraction\cite{Jeon_Kim_Jung_Sung_2022, Liu_Zheng_Wang_Zhang_2021, Gürtler_Dieter_Martius_2021}. 
These early works demonstrated that learning cooperation at the level of abstraction notably accelerates the learning process compared to flat methods. 
Drawing inspiration from hierarchical reinforcement learning, this study introduces a novel hierarchical framework that integrates the generation of topology by representing temporal abstraction. 
Topology generation is not limited to DAG generation methods. 
The latest continuous optimization frameworks for DAG learning include dag-notear\cite{Zheng_Aragam_Ravikumar_Xing_2018}, dag-nocurl\cite{Yu_Gao_Yin_Ji_2021}, and others\cite{Yu_Chen_Gao_Yu_2019, Lachapelle_Brouillard_Tristan_Lacoste-Julien_2019}.


Currently, in the field of multi-agent reinforcement learning (MARL), three main Centralized Training and Decentralized Execution (CTDE)\cite{Foerster_Assael_Freitas_Whiteson_2016, Gupta_Egorov_Kochenderfer_2017} paradigms are predominantly followed: multi-agent policy gradient methods\cite{Yu_Velu_Vinitsky_Wang_Bayen_Wu_2021}, value decomposition methods\cite{ Rashid_Samvelyan_Witt_Farquhar_Foerster_Whiteson_2018, Son_Kim_Kang_Hostallero_Ye_2019, Wang_Ren_Liu_Yu_Zhang_2020}, and actor-critic methods\cite{Yuan_Wang_Wang_Zhang_Chen_Guan_Zhang_Zhang_Yu}. 
These approaches typically assume independence among agents' policies and update all agents simultaneously. 
However, this simultaneous updating introduces non-stationarity issues\cite{Hernandez-Leal_Kaisers_Baarslag_Cote_2017}, especially in complex tasks where agents need to adapt to dynamic environments influenced by other agents' policies.

In contrast, algorithms employing sequential updates allow agents to perceive changes made by preceding agents, providing a more nuanced perspective on inter-agent interactions\cite{Gemp_Chen_McWilliams_2022}. 
Methods such as Rollout and Policy Iteration for a Single Agent (RPISA)\cite{Kakade_Langford_2002}, Heterogeneous Proximal Policy Optimization (HAPPO)\cite{Kuba_Chen_Wen_Wen_Sun_Wang_Yang_2021}, and Agent-by-agent Policy Optimization (A2PO)\cite{Wang_Tian_Wan_Wen_Wang_Zhang_2023} explore this sequential updating approach. 
RPISA effectively addresses non-stationarity by transforming multi-agent problems into stable single-agent ones, leading to more robust performance. 
However, it sacrifices sample inefficiency as only a single rollout sample is used to update a single-agent policy.
While HAPPO improves sample efficiency and exhibits monotonic improvement in joint policies, it does not theoretically guarantee policy improvement for individual agents.
Although A2PO ensures monotonic improvement for both joint and individual policies with a single rollout, its training time of sequential updates increases with the number of agents, resulting in low efficiency for joint policy training in large-scale agent clusters.

\section{Conclusion}
In conclusion, we introduce the B2MAPO scheme with the off-policy correction method, which theoretically guarantees the monotonic incrementally tightened bound. 
Then, a universal modulized plug-and-play B2MAPO hierarchical framework, which satisfies CTDE principle, is designed to conveniently integrate any MARL models to fully exploit and merge their merits. 
Based on B2MAPO framework, a deliberative implementation, called DAG-based B2MAPO algorithm is proposed. 
The upper layer employs PPO algorithm with attention mechanism to generate DAGs of agents and produce optimal batch sequence through topological sorting. 
Based on the theorem, the lower layer trains two joint policies in parallel and minimizes KL divergence between them periodically.  
The final derived joint policy achieves high efficiency while ensuring algorithm optimality. 
Experiments consistently show superior performance compared to baseline methods while significantly improving the efficiency during the training and execution phase. 
Overall, B2MAPO presents a promising direction in multi-agent decision-making, offering a principled approach that balances performance, efficiency, and adaptability.

\bibliography{m2559}

\begin{thebibliography}{45}
\providecommand{\natexlab}[1]{#1}
\providecommand{\url}[1]{\texttt{#1}}
\expandafter\ifx\csname urlstyle\endcsname\relax
  \providecommand{\doi}[1]{doi: #1}\else
  \providecommand{\doi}{doi: \begingroup \urlstyle{rm}\Url}\fi

\bibitem[Bernstein et~al.(2002)Bernstein, Givan, Immerman, and Zilberstein]{Bernstein_Givan_Immerman_Zilberstein_2002}
D.~S. Bernstein, R.~Givan, N.~Immerman, and S.~Zilberstein.
\newblock The complexity of decentralized control of markov decision processes.
\newblock \emph{Mathematics of Operations Research}, page 819–840, Nov 2002.
\newblock \doi{10.1287/moor.27.4.819.297}.
\newblock URL \url{http://dx.doi.org/10.1287/moor.27.4.819.297}.

\bibitem[Bertsekas(2021)]{Bertsekas_2021}
D.~Bertsekas.
\newblock Multiagent reinforcement learning: Rollout and policy iteration.
\newblock \emph{IEEE/CAA Journal of Automatica Sinica}, page 249–272, Feb 2021.
\newblock \doi{10.1109/jas.2021.1003814}.
\newblock URL \url{http://dx.doi.org/10.1109/jas.2021.1003814}.

\bibitem[Caprara et~al.(2000)Caprara, Toth, and Fischetti]{caprara2000algorithms}
A.~Caprara, P.~Toth, and M.~Fischetti.
\newblock Algorithms for the set covering problem.
\newblock \emph{Annals of Operations Research}, 98\penalty0 (1):\penalty0 353--371, 2000.

\bibitem[Cordeiro et~al.(2010)Cordeiro, Mounié, Perarnau, Trystram, Vincent, and Wagner]{Cordeiro_Mounié_Perarnau_Trystram_Vincent_Wagner_2010}
D.~Cordeiro, G.~Mounié, S.~Perarnau, D.~Trystram, J.-M. Vincent, and F.~Wagner.
\newblock Random graph generation for scheduling simulations.
\newblock \emph{Le Centre pour la Communication Scientifique Directe - HAL - Diderot,Le Centre pour la Communication Scientifique Directe - HAL - Diderot}, Mar 2010.

\bibitem[Dayan and Hinton(1992)]{Dayan_Hinton_1992}
P.~Dayan and G.~Hinton.
\newblock Feudal reinforcement learning.
\newblock \emph{Neural Information Processing Systems,Neural Information Processing Systems}, Nov 1992.

\bibitem[Demin et~al.(2021)Demin, Efimenko, Blinov, Komkova, and Rogov]{2021Multi}
V.~A. Demin, D.~B. Efimenko, D.~V. Blinov, D.~A. Komkova, and V.~R. Rogov.
\newblock Multi-agent approach to freight transportation using pooling technology.
\newblock In \emph{2021 Systems of Signals Generating and Processing in the Field of on Board Communications}, 2021.

\bibitem[Dwiel et~al.(2019)Dwiel, Candadai, Phielipp, and Bansal]{Dwiel_Candadai_Phielipp_Bansal_2019}
Z.~Dwiel, M.~Candadai, M.~Phielipp, and A.~Bansal.
\newblock Hierarchical policy learning is sensitive to goal space design.
\newblock \emph{Cornell University - arXiv,Cornell University - arXiv}, May 2019.

\bibitem[Foerster et~al.(2016)Foerster, Assael, Freitas, and Whiteson]{Foerster_Assael_Freitas_Whiteson_2016}
J.~Foerster, Y.~Assael, N.~Freitas, and S.~Whiteson.
\newblock Learning to communicate with deep multi-agent reinforcement learning.
\newblock \emph{arXiv: Artificial Intelligence,arXiv: Artificial Intelligence}, May 2016.

\bibitem[Gemp et~al.(2022)Gemp, Chen, and McWilliams]{Gemp_Chen_McWilliams_2022}
I.~Gemp, C.~Chen, and B.~McWilliams.
\newblock The generalized eigenvalue problem as a nash equilibrium.
\newblock \emph{arXiv preprint arXiv:2206.04993}, 2022.

\bibitem[Gupta et~al.(2017)Gupta, Egorov, and Kochenderfer]{Gupta_Egorov_Kochenderfer_2017}
J.~K. Gupta, M.~Egorov, and M.~Kochenderfer.
\newblock Cooperative multi-agent control using deep reinforcement learning.
\newblock In \emph{Autonomous Agents and Multiagent Systems: AAMAS 2017 Workshops, Best Papers, S{\~a}o Paulo, Brazil, May 8-12, 2017, Revised Selected Papers 16}, pages 66--83. Springer, 2017.

\bibitem[Gürtler et~al.(2021)Gürtler, Dieter, and Martius]{Gürtler_Dieter_Martius_2021}
N.~Gürtler, B.~Dieter, and G.~Martius.
\newblock Hierarchical reinforcement learning with timed subgoals.
\newblock \emph{Neural Information Processing Systems, Neural Information Processing Systems}, Dec 2021.

\bibitem[Han et~al.(2023)Han, Feng, Zhou, and Li]{han2023sample}
J.~Han, M.~Feng, W.~Zhou, and H.~Li.
\newblock Sample efficient reinforcement learning with double importance sampling weight clipping.
\newblock In \emph{2023 IEEE Conference on Games (CoG)}, pages 1--8. IEEE, 2023.

\bibitem[Han and Sung(2019)]{han2019dimension}
S.~Han and Y.~Sung.
\newblock Dimension-wise importance sampling weight clipping for sample-efficient reinforcement learning.
\newblock In \emph{International Conference on Machine Learning}, pages 2586--2595. PMLR, 2019.

\bibitem[Hernandez-Leal et~al.(2017)Hernandez-Leal, Kaisers, Baarslag, and Cote]{Hernandez-Leal_Kaisers_Baarslag_Cote_2017}
P.~Hernandez-Leal, M.~Kaisers, T.~Baarslag, and E.~Cote.
\newblock A survey of learning in multiagent environments: Dealing with non-stationarity.
\newblock \emph{arXiv: Multiagent Systems,arXiv: Multiagent Systems}, Jul 2017.

\bibitem[Hochba(1997)]{hochba1997approximation}
D.~S. Hochba.
\newblock Approximation algorithms for np-hard problems.
\newblock \emph{ACM Sigact News}, 28\penalty0 (2):\penalty0 40--52, 1997.

\bibitem[Jeon et~al.(2022)Jeon, Kim, Jung, and Sung]{Jeon_Kim_Jung_Sung_2022}
J.~Jeon, W.~Kim, W.~Jung, and Y.~Sung.
\newblock Maser: Multi-agent reinforcement learning with subgoals generated from experience replay buffer.
\newblock In \emph{International Conference on Machine Learning}, pages 10041--10052. PMLR, 2022.

\bibitem[Jiang et~al.(2021)Jiang, Li, Sun, and Zheng]{Jiang_Li_Sun_Zheng_2021}
Q.~Jiang, J.~Li, W.~Sun, and B.~Zheng.
\newblock Dynamic lane traffic signal control with group attention and multi-timescale reinforcement learning.
\newblock In \emph{Proceedings of the Thirtieth International Joint Conference on Artificial Intelligence}, Aug 2021.
\newblock \doi{10.24963/ijcai.2021/501}.
\newblock URL \url{http://dx.doi.org/10.24963/ijcai.2021/501}.

\bibitem[Kakade and Langford(2002)]{Kakade_Langford_2002}
S.~Kakade and J.~Langford.
\newblock Approximately optimal approximate reinforcement learning.
\newblock \emph{International Conference on Machine Learning, International Conference on Machine Learning}, Jul 2002.

\bibitem[Kuba et~al.(2021)Kuba, Chen, Wen, Wen, Sun, Wang, and Yang]{Kuba_Chen_Wen_Wen_Sun_Wang_Yang_2021}
J.~Kuba, R.~Chen, M.~Wen, Y.~Wen, F.~Sun, J.~Wang, and Y.~Yang.
\newblock Trust region policy optimisation in multi-agent reinforcement learning.
\newblock \emph{Cornell University - arXiv,Cornell University - arXiv}, Sep 2021.

\bibitem[Kurach et~al.(2019)Kurach, Raichuk, Stańczyk, Zając, Bachem, Espeholt, Riquelme, Vincent, Michalski, Bousquet, and Gelly]{Kurach_Raichuk_Stańczyk_Zając_Bachem_Espeholt_Riquelme_Vincent_Michalski_Bousquet_et}
K.~Kurach, A.~Raichuk, P.~Stańczyk, M.~Zając, O.~Bachem, L.~Espeholt, C.~Riquelme, D.~Vincent, M.~Michalski, O.~Bousquet, and S.~Gelly.
\newblock Google research football: A novel reinforcement learning environment.
\newblock \emph{arXiv: Learning,arXiv: Learning}, Jul 2019.

\bibitem[Lachapelle et~al.(2019)Lachapelle, Brouillard, Tristan, and Lacoste-Julien]{Lachapelle_Brouillard_Tristan_Lacoste-Julien_2019}
S.~Lachapelle, P.~Brouillard, D.~Tristan, and S.~Lacoste-Julien.
\newblock Gradient-based neural dag learning.
\newblock \emph{arXiv: Learning,arXiv: Learning}, Jun 2019.

\bibitem[Levy et~al.(2017)Levy, Konidaris, Platt, and Saenko]{Levy_Konidaris_Platt_Saenko_2017}
A.~Levy, G.~Konidaris, R.~Platt, and K.~Saenko.
\newblock Learning multi-level hierarchies with hindsight.
\newblock \emph{arXiv: Artificial Intelligence,arXiv: Artificial Intelligence}, Dec 2017.

\bibitem[Liu et~al.(2021)Liu, Zheng, Wang, and Zhang]{Liu_Zheng_Wang_Zhang_2021}
S.~Liu, L.~Zheng, J.~Wang, and C.~Zhang.
\newblock Learning subgoal representations with slow dynamics.
\newblock \emph{International Conference on Learning Representations, International Conference on Learning Representations}, May 2021.

\bibitem[Liu et~al.(2023)Liu, Luo, Yuan, Li, Jin, Chen, and Pan]{Liu_Luo_Yuan_Li_Jin_Chen_Pan}
Y.~Liu, G.~Luo, Q.~Yuan, J.~Li, L.~Jin, B.~Chen, and R.~Pan.
\newblock Gplight: grouped multi-agent reinforcement learning for large-scale traffic signal control.
\newblock In \emph{Proceedings of the Thirty-Second International Joint Conference on Artificial Intelligence}, pages 199--207, 2023.

\bibitem[Lowe et~al.(2017)Lowe, Wu, Tamar, Harb, Abbeel, and Mordatch]{Lowe_Wu_Tamar_Harb_Abbeel_Mordatch_2017}
R.~Lowe, Y.~Wu, A.~Tamar, J.~Harb, O.~Abbeel, and I.~Mordatch.
\newblock Multi-agent actor-critic for mixed cooperative-competitive environments.
\newblock \emph{Neural Information Processing Systems,Neural Information Processing Systems}, Jun 2017.

\bibitem[Munkres(2000)]{munkres2000topology}
J.~R. Munkres.
\newblock Topology, vol. 2, 2000.

\bibitem[Nachum et~al.(2018)Nachum, Gu, Lee, and Levine]{Nachum_Gu_Lee_Levine_2018}
O.~Nachum, S.~Gu, H.~Lee, and S.~Levine.
\newblock Data-efficient hierarchical reinforcement learning.
\newblock \emph{arXiv: Learning,arXiv: Learning}, May 2018.

\bibitem[Palanisamy(2019)]{2019Multi_2}
P.~Palanisamy.
\newblock Multi-agent deep reinforcement learning for connected autonomous driving.
\newblock In \emph{Neural Information Processing Systems (NeurIPS 2019), Machine Learning for Autonomous Driving Workshop}, 2019.

\bibitem[Palanisamy(2020)]{2019Multi}
P.~Palanisamy.
\newblock Multi-agent connected autonomous driving using deep reinforcement learning.
\newblock In \emph{2020 International Joint Conference on Neural Networks (IJCNN)}, pages 1--7. IEEE, 2020.

\bibitem[Rashid et~al.(2018)Rashid, Samvelyan, Witt, Farquhar, Foerster, and Whiteson]{Rashid_Samvelyan_Witt_Farquhar_Foerster_Whiteson_2018}
T.~Rashid, M.~Samvelyan, C.~Witt, G.~Farquhar, J.~Foerster, and S.~Whiteson.
\newblock Qmix: Monotonic value function factorisation for deep multi-agent reinforcement learning.
\newblock \emph{arXiv: Learning,arXiv: Learning}, Mar 2018.

\bibitem[Samvelyan et~al.(2019)Samvelyan, Rashid, Witt, Farquhar, Nardelli, Rudner, Hung, Torr, Foerster, and Whiteson]{Samvelyan_Rashid_Witt_Farquhar_Nardelli_Rudner_Hung_Torr_Foerster_Whiteson_2019}
M.~Samvelyan, T.~Rashid, C.~Witt, G.~Farquhar, N.~Nardelli, T.~Rudner, C.-M. Hung, P.~Torr, J.~Foerster, and S.~Whiteson.
\newblock The starcraft multi-agent challenge.
\newblock \emph{arXiv: Learning,arXiv: Learning}, Feb 2019.

\bibitem[Schaeffer et~al.(1993)Schaeffer, Lu, Szafron, and Lake]{schaeffer1993re}
J.~Schaeffer, P.~Lu, D.~Szafron, and R.~Lake.
\newblock A re-examination of brute-force search.
\newblock In \emph{Proceedings of the AAAI Fall Symposium on Games: Planning and Learning}, pages 51--58. AAAI Press Menlo Park, Calif., 1993.

\bibitem[Schulman et~al.(2015)Schulman, Moritz, Levine, Jordan, and Abbeel]{schulman2015high}
J.~Schulman, P.~Moritz, S.~Levine, M.~Jordan, and P.~Abbeel.
\newblock High-dimensional continuous control using generalized advantage estimation.
\newblock \emph{arXiv preprint arXiv:1506.02438}, 2015.

\bibitem[Schulman et~al.(2017)Schulman, Wolski, Dhariwal, Radford, and Klimov]{schulman2017proximal}
J.~Schulman, F.~Wolski, P.~Dhariwal, A.~Radford, and O.~Klimov.
\newblock Proximal policy optimization algorithms.
\newblock \emph{arXiv preprint arXiv:1707.06347}, 2017.

\bibitem[Son et~al.(2019)Son, Kim, Kang, Hostallero, and Ye]{Son_Kim_Kang_Hostallero_Ye_2019}
K.~Son, D.~Kim, W.~Kang, D.~Hostallero, and Y.~Ye.
\newblock Qtran: Learning to factorize with transformation for cooperative multi-agent reinforcement learning.
\newblock \emph{International Conference on Machine Learning, International Conference on Machine Learning}, May 2019.

\bibitem[Sutton et~al.(1999)Sutton, Precup, and Singh]{Sutton_Precup_Singh_1999}
R.~S. Sutton, D.~Precup, and S.~Singh.
\newblock Between mdps and semi-mdps: A framework for temporal abstraction in reinforcement learning.
\newblock \emph{Artificial Intelligence}, page 181–211, Aug 1999.
\newblock \doi{10.1016/s0004-3702(99)00052-1}.
\newblock URL \url{http://dx.doi.org/10.1016/s0004-3702(99)00052-1}.

\bibitem[Vaswani et~al.(2017)Vaswani, Shazeer, Parmar, Uszkoreit, Jones, Gomez, Kaiser, and Polosukhin]{vaswani2017attention}
A.~Vaswani, N.~Shazeer, N.~Parmar, J.~Uszkoreit, L.~Jones, A.~N. Gomez, {\L}.~Kaiser, and I.~Polosukhin.
\newblock Attention is all you need.
\newblock \emph{Advances in neural information processing systems}, 30, 2017.

\bibitem[Vezhnevets et~al.(2017)Vezhnevets, Osindero, Schaul, Heess, Jaderberg, Silver, and Kavukcuoglu]{Vezhnevets_Osindero_Schaul_Heess_Jaderberg_Silver_Kavukcuoglu_2017}
A.~Vezhnevets, S.~Osindero, T.~Schaul, N.~Heess, M.~Jaderberg, D.~Silver, and K.~Kavukcuoglu.
\newblock Feudal networks for hierarchical reinforcement learning.
\newblock \emph{arXiv: Artificial Intelligence,arXiv: Artificial Intelligence}, Mar 2017.

\bibitem[Wang et~al.(2020)Wang, Ren, Liu, Yu, and Zhang]{Wang_Ren_Liu_Yu_Zhang_2020}
J.~Wang, Z.~Ren, Z.~Liu, Y.~Yu, and C.~Zhang.
\newblock Qplex: Duplex dueling multi-agent q-learning.
\newblock \emph{Cornell University - arXiv,Cornell University - arXiv}, Aug 2020.

\bibitem[Wang et~al.(2023)Wang, Tian, Wan, Wen, Wang, and Zhang]{Wang_Tian_Wan_Wen_Wang_Zhang_2023}
X.~Wang, Z.~Tian, Z.~Wan, Y.~Wen, J.~Wang, and W.~Zhang.
\newblock Order matters: Agent-by-agent policy optimization.
\newblock \emph{arXiv preprint arXiv:2302.06205}, 2023.

\bibitem[Yu et~al.(2021{\natexlab{a}})Yu, Velu, Vinitsky, Wang, Bayen, and Wu]{Yu_Velu_Vinitsky_Wang_Bayen_Wu_2021}
C.~Yu, A.~Velu, E.~Vinitsky, Y.~Wang, A.~Bayen, and Y.~Wu.
\newblock The surprising effectiveness of mappo in cooperative, multi-agent games.
\newblock \emph{arXiv: Learning,arXiv: Learning}, Mar 2021{\natexlab{a}}.

\bibitem[Yu et~al.(2019)Yu, Chen, Gao, and Yu]{Yu_Chen_Gao_Yu_2019}
Y.~Yu, J.~Chen, T.~Gao, and M.~Yu.
\newblock Dag-gnn: Dag structure learning with graph neural networks.
\newblock \emph{Cornell University - arXiv,Cornell University - arXiv}, Apr 2019.

\bibitem[Yu et~al.(2021{\natexlab{b}})Yu, Gao, Yin, and Ji]{Yu_Gao_Yin_Ji_2021}
Y.~Yu, T.~Gao, N.~Yin, and Q.~Ji.
\newblock Dags with no curl: An efficient dag structure learning approach.
\newblock \emph{arXiv: Learning,arXiv: Learning}, Jun 2021{\natexlab{b}}.

\bibitem[Yuan et~al.(2022)Yuan, Wang, Wang, Zhang, Chen, Guan, Zhang, Zhang, and Yu]{Yuan_Wang_Wang_Zhang_Chen_Guan_Zhang_Zhang_Yu}
L.~Yuan, C.~Wang, J.~Wang, F.~Zhang, F.~Chen, C.~Guan, Z.~Zhang, C.~Zhang, and Y.~Yu.
\newblock Multi-agent concentrative coordination with decentralized task representation.
\newblock In \emph{IJCAI}, pages 599--605, 2022.

\bibitem[Zheng et~al.(2018)Zheng, Aragam, Ravikumar, and Xing]{Zheng_Aragam_Ravikumar_Xing_2018}
X.~Zheng, B.~Aragam, P.~Ravikumar, and E.~Xing.
\newblock Dags with no tears: Continuous optimization for structure learning.
\newblock \emph{Neural Information Processing Systems,Neural Information Processing Systems}, Jan 2018.

\end{thebibliography}






\newpage
\appendix
\onecolumn
\section{Appendix}
\textbf{Corollary 1} For all $s$, $D_{TV}(\boldsymbol{\pi}(\cdot|s)\|\boldsymbol{\hat{\pi}}(\cdot|s))\leq\sum_{i=1}^nD_{TV}(\pi^i(\cdot|s)\|\hat{\pi}^i(\cdot|s))$, where $D_{TV}(p\|q)$ is the total variation distance between distributions $p$ and $q$.

$Proof.$ We denote $\pi( \cdot |s) $ as $\pi( \cdot ) $ for brevity.

$$
\begin{aligned}
&D_{TV}(\boldsymbol{\pi}(\cdot|s)\|\boldsymbol{\hat{\pi}}(\cdot|s)) \\
=&\frac12\sum_{a^1,a^2,...,a^n}\left|\prod_{i=1}^n\pi^i(a^i)-\prod_{i=1}^n\hat{\pi}^i(a^i)\right| \\
=&\frac12\sum_{a^1,a^2,...,a^n}\left|\prod_{i=1}^n\pi^i(a^i)-\pi^1(a^1)\prod_{i=2}^n\hat{\pi}^i(a^i)+\pi^1(a^1)\prod_{i=2}^n\hat{\pi}^i(a^i)-\prod_{i=1}^n\hat{\pi}^i(a^i)\right| \\
\leq&\frac12\sum_{a^1}\left|\pi^1(a^1)\right|\sum_{a^2,...,a^n}\left|\prod_{i=2}^n\pi^i(a^i)-\prod_{i=2}^n\hat{\pi}^i(a^i)\right|+\frac12\sum_{a^1}\left|\pi^1(a^1)-\hat{\pi}^1(a^1)\right|\sum_{a^2,...,a^n}\left|\prod_{i=2}^n\pi^i(a^i)\right| \\
=&\frac12\sum_{a^2,\ldots,a^n}\left|\prod_{i=2}^n\pi^i(a^i)-\prod_{i=2}^n\hat{\pi}^i(a^i)\right|+\frac12\sum_{a^1}\left|\pi^1(a^1)-\hat{\pi}^1(a^1)\right| \\
& \cdots\\
\leq&\frac12\sum_{i=1}^n\sum_{a^i}|\pi^i(a^i)-\hat{\pi}^i(a^i)| \\
=&\sum_{i=1}^nD_{TV}(\pi^i(\cdot|s)\|\hat{\pi}^i(\cdot|s))
\end{aligned}
$$

\textbf{Lemma 1} Given any joint policies $\boldsymbol{\hat{\pi}}$ and $\boldsymbol{\pi}$, we define $D_{TV}^{max}(\pi\|\hat{\pi}) = \max_{s}{D_{TV}(\pi(\cdot|s)\|\hat{\pi}}(\cdot|s)) $, then the following inequality holds:

$$
|\mathbb{E}_{\boldsymbol{a}\sim\hat{\boldsymbol{\pi}}}\left[A^{\boldsymbol{\pi}}(s,\boldsymbol{a})\right]|\leq2\epsilon\sum_{i=1}^{n}\alpha^{i}\:
$$
 where $\alpha^i=D_{TV}^{max}(\hat{\pi}^i\|\pi^i)$ and $\epsilon=\max_{s,\boldsymbol{a}}|A^{\boldsymbol{\pi}}(s,\boldsymbol{a})|.$

$Proof.$ Note that $\mathbb{E}_{\boldsymbol{a\sim\pi}}[A^{\boldsymbol{\pi}}(s,\boldsymbol{a})]=0.$ We have

$$
\begin{aligned}
|\mathbb{E}_{\boldsymbol{a}\sim\hat{\boldsymbol{\pi}}}\left[A^{\boldsymbol{\pi}}(s,\boldsymbol{a})\right]|& =|\mathbb{E}_{\hat{\boldsymbol{a}}\sim\hat{\boldsymbol{\pi}}}\left[A^{\boldsymbol{\pi}}(s,\hat{\boldsymbol{a}})\right]-\mathbb{E}_{\boldsymbol{a}\sim\boldsymbol{\pi}}\left[A^{\boldsymbol{\pi}}(s,\boldsymbol{a})\right]|  \\
&=\left|\mathbb{E}_{(\hat{\boldsymbol{a}},\boldsymbol{a})\sim(\hat{\boldsymbol{\pi}},\boldsymbol{\pi})}\left[A^{\boldsymbol{\pi}}(s,\hat{\boldsymbol{a}})-A^{\boldsymbol{\pi}}(s,\boldsymbol{a})\right]\right| \\
&=\left|Pr(\hat{\boldsymbol{a}}\neq\boldsymbol{a}|s)\mathbb{E}_{(\hat{\boldsymbol{a}},\boldsymbol{a})\sim(\hat{\boldsymbol{\pi}},\boldsymbol{\pi})}\left[A^{\boldsymbol{\pi}}(s,\hat{\boldsymbol{a}})-A^{\boldsymbol{\pi}}(s,\boldsymbol{a})\right]\right| \\
&\leq\sum_{i=1}^n\alpha^i\mathbb{E}_{(\hat{\boldsymbol{a}},\boldsymbol{a})\sim(\hat{\boldsymbol{\pi}},\boldsymbol{\pi})}\left[|A^{\boldsymbol{\pi}}(s,\hat{\boldsymbol{a}})-A^{\boldsymbol{\pi}}(s,\boldsymbol{a})|\right] \\
&\leq\sum_{i=1}^n\alpha^i\cdot2\max_{s,\boldsymbol{a}}|A^{\boldsymbol{\pi}}(s,\boldsymbol{a})|
\end{aligned}
$$

\textbf{Lemma 2 (Multi-agent Advantage Discrepancy Lemma)}. Given any joint policies $\boldsymbol{\pi}^1,\boldsymbol{\pi}^2$ and $\boldsymbol{\pi}^3$ the following inequality holds:

$$
\begin{aligned}&\left|\mathbb{E}_{(s_t,\boldsymbol{a}_t)\sim(Pr^{\boldsymbol{\pi}^2},\boldsymbol{\pi}^2)}\left[A^{\boldsymbol{\pi}^1}\right]-\mathbb{E}_{(s_t,\hat{\boldsymbol{a}}_t)\sim(Pr^{\boldsymbol{\pi}^3},\boldsymbol{\pi}^2)}\left[A^{\boldsymbol{\pi}^1}\right]\right|\\\leq&4\epsilon^{\boldsymbol{\pi}^1}\cdot D_{TV}^{max}(\boldsymbol{\pi}^1\|\boldsymbol{\pi}^2)\cdot(1-(1-D_{TV}^{max}(\boldsymbol{\pi}^2\|\boldsymbol{\pi}^3))^t)\:\end{aligned}
$$
 where $\epsilon^{\boldsymbol{\pi}^1}=\max_{s,\boldsymbol{a}}\|A^{\boldsymbol{\pi}^1}(s,\boldsymbol{a})\|$ and we denote $A(s,\boldsymbol{a})$ as $A$ for brevity.
 
$Proof$. Let $n_t$ represent the times $\boldsymbol{a}\neq\boldsymbol{\hat{a}}\left(\boldsymbol{\pi}^1 \text{ disagrees with } \boldsymbol{\pi}^3\right)$ before timestamp $t$.
$$
\begin{aligned}
&\left|\mathbb{E}_{(s_t,\boldsymbol{a}_t)\sim(Pr^{\boldsymbol{\pi}^2},\boldsymbol{\pi}^2)}\left[A^{\boldsymbol{\pi}^1}\right]-\mathbb{E}_{(s_t,\hat{\boldsymbol{a}}_t)\sim(Pr^{\boldsymbol{\pi}^3},\boldsymbol{\pi}^2)}\left[A^{\boldsymbol{\pi}^1}\right]\right| \\
=&Pr(n_t>0)\cdot\left|\mathbb{E}_{(s_t,\boldsymbol{a}_t)\sim(Pr^{\boldsymbol{\pi}^2},\boldsymbol{\pi}^2)|n_t>0}\left[A^{\boldsymbol{\pi}^1}\right]-\mathbb{E}_{(s_t,\hat{\boldsymbol{a}}_t)\sim(Pr^{\boldsymbol{\pi}^3},\boldsymbol{\pi}^2)|n_t>0}\left[A^{\boldsymbol{\pi}^1}\right]\right| \\
\stackrel{(a)}{=}&(1-Pr(n_{t}=0))\cdot E \\
\leq&(1-\prod_{k=1}^{t}Pr(\boldsymbol{a}_{k}=\hat{\boldsymbol{a}}_{k}|\boldsymbol{a}_{k}\sim\boldsymbol{\pi}^{2}(\cdot|s_{k}),\hat{\boldsymbol{a}}_{k}\sim\boldsymbol{\pi}^{3}(\cdot|s_{k})))\cdot E \\
\stackrel{(b)}{\leq}&(1-\prod_{k=1}^{t}(1-D_{TV}^{max}(\boldsymbol{\pi}^{2}\|\boldsymbol{\pi}^{3})))\cdot E \\
=&(1-(1-D_{TV}^{max}(\boldsymbol{\pi}^{2}\|\boldsymbol{\pi}^{3}))^{t})\cdot E \\
\leq&(1-(1-D_{TV}^{max}(\boldsymbol{\pi}^{2}\|\boldsymbol{\pi}^{3}))^{t})\cdot2\cdot2\cdot D_{TV}^{max}(\boldsymbol{\pi}^{1}\|\boldsymbol{\pi}^{2})\cdot\epsilon^{\boldsymbol{\pi}^{1}} \\
=&4\epsilon^{\boldsymbol{\pi}^{1}}\cdot D_{TV}^{max}(\boldsymbol{\pi}^{1}\|\boldsymbol{\pi}^{2})\cdot(1-(1-D_{TV}^{max}(\boldsymbol{\pi}^{2}\|\boldsymbol{\pi}^{3}))^{t})
\end{aligned}
$$

In (a), we denote $|\mathbb{E}_{(s_t,\boldsymbol{a}_t)\sim(Pr^{\boldsymbol{\pi}^2},\boldsymbol{\pi}^2)|n_t>0}[A^{\boldsymbol{\pi}^1}]-\mathbb{E}_{(s_t,\hat{\boldsymbol{a}}_t)\sim(Pr^{\boldsymbol{\pi}^3},\boldsymbol{\pi}^2)|n_t>0}[A^{\boldsymbol{\pi}^1}]|$ as $E$ for brevity.

We provide a useful equation of the normalized discounted state visitation distribution here.

\textbf{Proposition 1}
$$
\begin{aligned}
\mathbb{E}_{(s,\boldsymbol{a})\sim(d^{\boldsymbol{\pi}^1},\boldsymbol{\pi}^2)}\left[f(s,\boldsymbol{a})\right]& =(1-\gamma)\sum_{s}\sum_{t=0}^{\infty}\gamma^{t}Pr(s_{t}=s|\boldsymbol{\pi}^{1})\sum_{\boldsymbol{a}}\boldsymbol{\pi}^{2}(\boldsymbol{a}|s)f(s,\boldsymbol{a})  \\
&=(1-\gamma)\sum_{t=0}^{\infty}\gamma^t\sum_{s}Pr(s_t=s|\boldsymbol{\pi}^1)\sum_{\boldsymbol{a}}\boldsymbol{\pi}^2(\boldsymbol{a}|s)f(s,\boldsymbol{a}) \\
&=(1-\gamma)\sum_{t=0}^\infty\gamma^t\mathbb{E}_{(s_t,\boldsymbol{a}_t)\sim(Pr^{\boldsymbol{\pi}^1},\boldsymbol{\pi}^2)}[f(s_t,\boldsymbol{a}_t)]
\end{aligned}
$$
\subsection{B2MAPO}
\textbf{Theorem 1.1 (Single Batch Monotonic Bound of B2MAPO) }

For batch $b_k$, let $\alpha^{b_k}=D_{TV}^{\max}(\pi^{b_k}\|\hat{\pi}^{b_k}), \epsilon = \max_{b_k}\epsilon^{b_k}, \epsilon^{b_k}=\max_{s,\boldsymbol{a}}|A^{\hat{\boldsymbol{\pi}}^{b_{k-1}}}(s,a)|,~\xi^{b_k}=$
    $\max_{s,\boldsymbol{a}}|A^{\boldsymbol{\pi},\hat{\boldsymbol{\pi}}^{b_{k-1}}}(s,a)-A^{\hat{\boldsymbol{\pi}}^{b_{k-1}}}(s,\boldsymbol{a})|$, then we have:

$$
\begin{aligned}
&\left|\mathcal{J}(\hat{\boldsymbol{\pi}}^{b_k})-\mathcal{L}_{\hat{\boldsymbol{\pi}}^{b_{k-1}}}(\hat{\boldsymbol{\pi}}^{b_k})\right| \leq\\
&4\epsilon^{b_k}\alpha^{b_k}\big(\frac{1}{1-\gamma}-\frac{1}{1-\gamma(1-\sum_{{b_j}\in(B_k\cup\{b_k\})}\alpha^{b_j}\big)}\big)+\frac{\xi^{b_k}}{1-\gamma} \\
&\leq\frac{4\gamma\epsilon^{b_k}}{(1-\gamma)^2}\big(\alpha^{b_k}\sum_{{b_j}\in(B_k\cup\{b_k\})}\alpha^{b_j}\big)+\frac{\xi^{b_k}}{1-\gamma}\:. 
\end{aligned}
$$

$Proof$.

$$
\begin{aligned}
&\left|\mathcal{J}(\hat{\boldsymbol{\pi}}^{b_k})-\mathcal{J}(\hat{\boldsymbol{\pi}}^{b_{k-1}})-\frac1{1-\gamma}\mathbb{E}_{(s,\boldsymbol{a})\sim(d^{\boldsymbol{\pi}},\hat{\boldsymbol{\pi}}^{b_k})}\left[A^{\boldsymbol{\pi},\hat{\boldsymbol{\pi}}^{b_{k-1}}}\right]\right| \\
&=\frac{1}{1-\gamma}\left|\mathbb{E}_{(s,\boldsymbol{a})\sim(d^{\hat{\boldsymbol{\pi}}^{b_k}},\hat{\boldsymbol{\pi}}^{b_k})}\left[A^{\hat{\boldsymbol{\pi}}^{{b_{k-1}}}}\right]-\mathbb{E}_{(s,\boldsymbol{a})\sim(d^{\boldsymbol{\pi}},\hat{\boldsymbol{\pi}}^{b_k})}\left[A^{\boldsymbol{\pi},\hat{\boldsymbol{\pi}}^{{b_{k-1}}}}\right]\right| \\
&\leq\frac{1}{1-\gamma}\left|\mathbb{E}_{(s,\boldsymbol{a})\sim(d^{\hat{\boldsymbol{\pi}}^{b_k}},\hat{\boldsymbol{\pi}}^{b_k})}\left[A^{\hat{\boldsymbol{\pi}}^{{b_{k-1}}}}\right]-\mathbb{E}_{(s,\boldsymbol{a})\sim(d^{\boldsymbol{\pi}},\hat{\boldsymbol{\pi}}^{b_k})}\left[A^{\hat{\boldsymbol{\pi}}^{{b_{k-1}}}}\right]\right| \\
&+\frac{1}{1-\gamma}\left|\mathbb{E}_{(s,\boldsymbol{a})\sim(d^{\boldsymbol{\pi}},\hat{\boldsymbol{\pi}}^{b_k})}\left[A^{\hat{\boldsymbol{\pi}}^{{b_{k-1}}}}\right]-\mathbb{E}_{(s,\boldsymbol{a})\sim(d^{\boldsymbol{\pi}},\hat{\boldsymbol{\pi}}^{b_k})}\left[A^{\boldsymbol{\pi},\hat{\boldsymbol{\pi}}^{{b_{k-1}}}}\right]\right| \\
&\leq4\epsilon^{b_{k}}\alpha^{{b_k}}\sum_{t=0}^{\infty}\gamma^{t}(1-(1-\sum_{j\in(B_k\cup\{{b_k}\})}\alpha^{j})^{t})+\frac{1}{1-\gamma}\mathbb{E}_{(s,\boldsymbol{a})\sim(d^{\boldsymbol{\pi}},\hat{\boldsymbol{\pi}}^{{b_k}})}\left[\left|A^{\hat{\boldsymbol{\pi}}^{b_{k-1}}}-A^{\boldsymbol{\pi},\hat{\boldsymbol{\pi}}^{{b_{k-1}}}}\right|\right] \\
&\leq4\epsilon^{{b_{k}}}\alpha^{{b_k}}(\frac{1}{1-\gamma}-\frac{1}{1-\gamma(1-\sum_{j\in(B_k\cup\{{b_k}\})}\alpha^{j})})+\frac{1}{1-\gamma}\xi^{b_k}
\end{aligned}
$$

\textbf{Theorem 1.2 (Joint Monotonic Bound of B2MAPO) }

For each  batch $b_k \in B$, let $\epsilon^{b_k}=\max_{s,\boldsymbol{a}}|A^{\hat{\boldsymbol{\pi}}^{b_{k-1}}}((s,\boldsymbol{a})|,
\alpha^{b_k}=D_{TV}^{\max}(\pi^{b_k}\|\hat{\pi}^{b_k}),\xi^{b_k}=\max_{s,\boldsymbol{a}}|A^{\boldsymbol{\pi},\hat{\boldsymbol{\pi}}^{b_{k-1}}}(s,\boldsymbol{a})-A^{\hat{\boldsymbol{\pi}}^{b_{k-1}}}(s,\boldsymbol{a})|$, and $\epsilon=\max_{b_k}\epsilon^{b_k}$,then we have:
$$
|\mathcal{J}(\hat{\boldsymbol{\pi}})-\mathcal{G}_{\boldsymbol{\pi}}(\hat{\boldsymbol{\pi}})| \leq
\frac{4\gamma\epsilon}{(1-\gamma)^2}\sum_{k=1}^{|B|}\left(\alpha^{b_k}\sum_{j\in(B_k\cup\{b_k\})}\alpha^{b_j}\right)+\frac{\sum_{k=1}^{|B|}\xi^{b_k}}{1-\gamma}. \\
$$

$Proof.$
$$
\begin{aligned}
&|\mathcal{J}(\hat{\boldsymbol{\pi}})-\mathcal{G}_{\boldsymbol{\pi}}(\hat{\boldsymbol{\pi}})| \\
&=\left|\mathcal{J}(\hat{\boldsymbol{\pi}})-\mathcal{J}(\boldsymbol{\pi})-\sum_{k=1}^{|B|}\mathbb{E}_{(s,\boldsymbol{a})
\sim(d^{\boldsymbol{\pi}},{\hat{\boldsymbol{\pi}}}^{b_k})}\left[A^{\boldsymbol{\pi},\hat{\boldsymbol{\pi}}^{b_{k-1}}}(s,\boldsymbol{a})\right]\right| \\
&=\left|\mathcal{J}(\hat{\boldsymbol{\pi}}^{b_m})-\mathcal{J}(\hat{\boldsymbol{\pi}}^{b_{m-1}})+\cdots+\mathcal{J}(\hat{\boldsymbol{\pi}}^{b_1})-\mathcal{J}(\boldsymbol{\pi})-\frac1{1-\gamma}\sum_{k=1}^{|B|}\mathbb{E}_{(s,\boldsymbol{a})\sim(d^{\boldsymbol{\pi}},\hat{\boldsymbol{\pi}}^{b_k})}\left[A^{\boldsymbol{\pi},\boldsymbol{\pi}^{b_{k-1}}}(s,\boldsymbol{a})\right] \right|. \\
&\leq\sum_{k=1}^{|B|}\left|\mathcal{J}(\hat{\boldsymbol{\pi}}^{b_k})-\mathcal{J}(\hat{\pi}^{b_{k-1}})-\frac1{1-\gamma}\mathbb{E}_{(s,\boldsymbol{a})\sim(d^{\boldsymbol{\pi}},\hat{\boldsymbol{\pi}}^{b_k})}\left[A^{\boldsymbol{\pi},\hat{\boldsymbol{\pi}}^{b_{k-1}}}(s,\boldsymbol{a})\right]\right| \\
&\leq4\epsilon\sum_{k=1}^{|B|}\alpha^{b_k}\left(\frac{1}{1-\gamma}-\frac{1}{1-\gamma(1-\sum_{j\in(B_k\cup\{b_k\})}\alpha^{j})}\right)+\frac{\sum_{k=1}^{|B|}\xi^{b_k}}{1-\gamma} \\
&\leq\frac{4\gamma\epsilon}{(1-\gamma)^2}\sum_{b_k=1}^{|B|}\left(\alpha^{b_k}\sum_{j\in(B_k\cup\{b_k\})}\alpha^j\right)+\frac{\sum_{k=1}^{|B|}\xi^{b_k}}{1-\gamma}.
\end{aligned}
$$
\textbf{Theorem 1.3 (Incrementally Tightened Bound of B2MAPO) }
Given $(B,\prec)$, assume policy is updated in the batch sequence $B$, then:
$$
\begin{aligned}
&|\mathcal{J}(\hat{\boldsymbol{\pi}})-\mathcal{G}_{\boldsymbol{\pi}}(\hat{\boldsymbol{\pi}})| \\
&\begin{aligned}\leq\sum_{k=1}^{b-1}\left|\mathcal{J}(\hat{\boldsymbol{\pi}}^{b_k})-\mathcal{L}_{\hat{\boldsymbol{\pi}}^{b_{k-1}}}(\hat{\boldsymbol{\pi}}^{b_k})\right|+4\epsilon\sum_{k=b}^{|B|}\alpha^{b_k}\left(\frac{1}{1-\gamma}-\frac{1}{1-\gamma(1-\sum_{j\in(B_k\cup\{b_k\})}\alpha^{j})}\right)+\frac{\sum_{k=b}^{|B|}\xi^{b_k}}{1-\gamma}\end{aligned} \\
&\leq\sum_{k=1}^{b-2}\left|\mathcal{J}(\hat{\boldsymbol{\pi}}^{b_k})-\mathcal{L}_{\hat{\boldsymbol{\pi}}^{b_{k-1}}}(\hat{\boldsymbol{\pi}}^{b_k})\right|+4\epsilon\sum_{k=b-1}^{|B|}\alpha^{b_k}\left(\frac1{1-\gamma}-\frac1{1-\gamma(1-\sum_{j\in\{B_k\cup\{{b_k}\}\}}\alpha^j)}\right)+\frac{\sum_{k=b-1}^{|B|}\xi^{b_k}}{1-\gamma} \\
&\leq4\epsilon\sum_{k=1}^{|B|}\alpha^{b_k}\left(\frac{1}{1-\gamma}-\frac{1}{1-\gamma(1-\sum_{j\in(B_k\cup\{{b_k}\})}\alpha^{j})}\right)+\frac{\sum_{k=1}^{|B|}\xi^{{b_k}}}{1-\gamma}
\end{aligned}
$$
\textbf{Theorem 2} 
For any joint policy, $\boldsymbol{\pi}_{A2PO}$, produced by A2PO, there exists the equivalent joint policy $\boldsymbol{\pi}_{B2MAPO}$ , that can be acquired by B2MAPO scheme.

$Proof$.
Incrementally Tightened Bound of A2PO is less than or equal to that of B2MAPO :

$\begin{aligned}
4\epsilon\sum_{k=1}^{n}\alpha^{i}\left(\frac{1}{1-\gamma}-\frac{1}{1-\gamma(1-\sum_{j\in({e^i}\cup\{{i}\})}\alpha^{j})}\right)+\frac{\sum_{k=1}^{n}\xi^{{i}}}{1-\gamma}
\leq4\epsilon\sum_{k=1}^{|B|}\alpha^{b_k}\left(\frac{1}{1-\gamma}-\frac{1}{1-\gamma(1-\sum_{j\in(B_k\cup\{{b_k}\})}\alpha^{j})}\right)+\frac{\sum_{k=1}^{|B|}\xi^{{b_k}}}{1-\gamma}
\end{aligned}$
 
And when $|B|=n$, the Bound of A2PO is equal to that of B2MAPO, so for any joint policy, $\boldsymbol{\pi}_{A2PO}$, produced by A2PO, there exists the equivalent joint policy, $\boldsymbol{\pi}_{B2MAPO}$ , that can be acquired by B2MAPO scheme.

\textbf{Theorem 3} For any joint policy $\boldsymbol{\pi}$ produced by B2MAPO scheme, there exists a joint policy $\boldsymbol{\pi}_{ind}$ acquired by simultaneous policy optimization scheme that does not involve dependency among agents, such that $V_{\boldsymbol{\pi}}(s) = V_{\boldsymbol{\pi}_{ind}}(s)$ for any state $s \in S$.

$Proof$. For a joint policy $\boldsymbol{\pi}$ produced by B2MAPO scheme, let $\max_{\boldsymbol{a}}Q_{\boldsymbol{\pi}} =A$ and $\min_{\boldsymbol{a}}Q_{\boldsymbol{\pi}}=B$, we have $A\leq V_{\boldsymbol{\pi}}( s) \leq B$. Then, we can construct the following joint policy $\boldsymbol{\pi}_{ind}$ follows simultaneous policy optimization scheme:
$$
\boldsymbol{\pi}_{ind}=\prod_{i=1}^n\pi_i=\prod_{i=1}^n\mathbf{1}[a_i=\arg\max Q_{\boldsymbol{\pi}}[i]].
$$

For such a joint policy $\boldsymbol{\pi}_{ind}$, we have $\sum_{\boldsymbol{a}}\boldsymbol{\pi}_{ind}Q_{\boldsymbol{\pi}}=A$. Similarly, we can also construct another joint policy, such that $\sum_{\boldsymbol{a}}\boldsymbol{\pi}_{ind}Q_{\boldsymbol{\pi}}=B.$ Based on the generalized intermediate value theorem \cite{munkres2000topology}, We can have that for any joint policy $\boldsymbol{\pi}$, there exists an joint policy $\boldsymbol{\pi}_{ind}$ such that:

$$
V_{\boldsymbol{\pi}}=\sum_{\boldsymbol{a}}\boldsymbol{\pi}Q_{\boldsymbol{\pi}}=\sum_{\boldsymbol{a}}\boldsymbol{\pi}_{ind}Q_{\boldsymbol{\pi}}=\mathbb{E}_{\boldsymbol{a}_{t}\sim\boldsymbol{\pi}_{ind}}[Q_{\boldsymbol{\pi}}].
$$

Thus, we can have:

$$
\begin{aligned}
V_{\boldsymbol{\pi}}(s_{t}) =&\mathbb{E}_{\boldsymbol{a}_{t}\sim\boldsymbol{\pi}_{ind}}[Q_{\boldsymbol{\pi}(s_{t},\boldsymbol{a}_{t})}] \\
=&\mathbb{E}_{\boldsymbol{a}_t\sim\boldsymbol{\pi}_{ind},s_{t+1}\sim P}[r(s_t,\boldsymbol{a}_t)+\gamma V_{\boldsymbol{\pi}}(s_{t+1})] \\
=&\mathbb{E}_{(\boldsymbol{a}_t,\boldsymbol{a}_{t+1})\sim \boldsymbol{\pi}_{ind},s_{t+1}\sim P}[r(s_t,a_t)+\gamma Q_{\boldsymbol{\pi}(s_t,a_t)]} \\
&\vdots\\
=&\mathbb{E}_{\boldsymbol{a}_{t:\infty}\sim\boldsymbol{\pi}_{ind},s_{t:\infty}\sim P}[r(s_t,a_t)+\gamma r(s_{t+1},a_{t+1})+\cdots] \\
=&V_{\boldsymbol{\pi}_{ind}}(s_t)
\end{aligned}
$$
 which concludes the proof.
\subsection{MAPPO}
For MAPPO, $\mathcal{L}_{\boldsymbol{\pi}}(\hat{\boldsymbol{\pi}})=\sum_{i=1}^n\mathcal{J}(\boldsymbol{\pi})+\frac1{1-\gamma}[\mathbb{E}_{(s,\boldsymbol{a})\sim(d^{\boldsymbol{\pi}},\boldsymbol{\pi})}[\frac{\hat{\pi}^i}{\pi^i}A^{\boldsymbol{\pi}}]].$ We first prove that 
is bounded. For agent $i,\mathcal{J}(\hat{\pi})-\mathcal{J}(\pi)-\frac1{1-\gamma}[\mathbb{E}_{(s,\boldsymbol{a})\sim(d^{\boldsymbol{\pi}},\boldsymbol{\pi})}[\frac{\hat{\pi}^i}{\pi^i}A^{\boldsymbol{\pi}}]]$ is bounded.

$$\begin{aligned}
&\left|\mathcal{J}(\hat{\boldsymbol{\pi}})-\mathcal{J}(\boldsymbol{\pi})-\frac1{1-\gamma}\left[\mathbb{E}_{(s,\boldsymbol{a})\sim(d^{\boldsymbol{\pi}},\boldsymbol{\pi})}\left[\frac{\hat{\pi}^i}{\pi^i}A^{\boldsymbol{\pi}}\right]\right]\right| \\
&= \frac1{1-\gamma}\left|\mathbb{E}_{(s,\boldsymbol{a})\sim(d^{\hat{\boldsymbol{\pi}}},\hat{\boldsymbol{\pi}})}\left[A^{\boldsymbol{\pi}}\right]-\mathbb{E}_{(s,\boldsymbol{a})\sim(d^{\boldsymbol{\pi}},\boldsymbol{\pi})}\left[\frac{\hat{\pi}^{i}}{\pi^{i}}A^{\boldsymbol{\pi}}\right]\right|  \\
&= \sum_{t=0}^{\infty}\gamma^{t}\left|\mathbb{E}_{(s_{t},\boldsymbol{a}_{t})\sim(Pr^{\hat{\boldsymbol{\pi}}},\hat{\boldsymbol{\pi}})}A^{\boldsymbol{\pi}}-\mathbb{E}_{(s_{t},\boldsymbol{a}_{t})\sim(Pr^{\boldsymbol{\pi}},\boldsymbol{\pi})}\left[\frac{\hat{\pi}^{i}}{\pi^{i}}A^{\pi}\right]\right|  \\
&\leq \sum_{t=0}^\infty2\gamma^t\left(\left(\sum_{j=1}^n\alpha^j\right)\cdot\epsilon^{\boldsymbol{\pi}}+\alpha^i\cdot\epsilon^{\boldsymbol{\pi}}\right)  \\
&= \frac{2\epsilon^{\boldsymbol{\pi}}}{1-\gamma}\left(\alpha^i+\sum_{j=1}^n\alpha^j\right) 
\end{aligned}$$
Sum the bounds for all agents and take the average, we get
$$\begin{aligned}&\left|\mathcal{J}(\hat{\boldsymbol{\pi}})-\mathcal{J}(\boldsymbol{\pi})-\frac{1}{n}\frac{1}{1-\gamma}\sum_{i=1}^{n}\left[\mathbb{E}_{(s,\boldsymbol{a})\sim(d^{\boldsymbol{\pi}},\boldsymbol{\pi})}\left[\frac{\hat{\pi}^{i}}{\pi^{i}}A^{\boldsymbol{\pi}}\right]\right]\right|\\\leq&\frac{2\epsilon^{\boldsymbol{\pi}}}{1-\gamma}\frac{n+1}{n}\sum_{j=1}^{n}\alpha^{j}\end{aligned}$$
Finally, the monotonic bound for MAPPO is
$$\begin{aligned}
&\left|\mathcal{J}(\hat{\boldsymbol{\pi}})-\mathcal{J}(\boldsymbol{\pi})-\frac{1}{1-\gamma}\sum_{i=1}^{n}\left[\mathbb{E}_{(s,\boldsymbol{a})\sim(d\boldsymbol{\pi},\boldsymbol{\pi})}\left[\frac{\hat{\pi}^{i}}{\pi^{i}}A^{\boldsymbol{\pi}}\right]\right]\right| \\
\leq&\left|\mathcal{J}(\hat{\boldsymbol{\pi}})-\mathcal{J}(\boldsymbol{\pi})-\frac1n\frac1{1-\gamma}\sum_{i=1}^n\left[\mathbb{E}_{(s,\boldsymbol{a})\sim(d^{\boldsymbol{\pi}},\boldsymbol{\pi})}\left[\frac{\hat{{\pi}}^i}{\pi^i}A^{\boldsymbol{\pi}}\right]\right]\right.  \\
&+\frac{n-1}{n}\frac{1}{1-\gamma}\left|\sum_{i=1}^{n}\left[\mathbb{E}_{(s,\boldsymbol{a})\sim(d^{\boldsymbol{\pi}},\boldsymbol{\pi})}\left[\frac{\hat{\pi}^{i}}{\pi^{i}}A^{\boldsymbol{\pi}}\right]\right]\right| \\
\leq &\frac{2\epsilon^{\boldsymbol{\pi}}}{1-\gamma}\frac{n+1}{n}\sum_{j=1}^{n}\alpha^{j}+\frac{n-1}{n}\sum_{i=1}^{n}\frac{1}{1-\gamma}\alpha^{i}\cdot2\epsilon^{\boldsymbol{\pi}}  \\
= &\begin{aligned}\frac{4\epsilon^{\boldsymbol{\pi}}}{1-\gamma}\sum_{i=1}^{n}\alpha^{i}\end{aligned} 
\end{aligned}$$
\subsection{HAPPO}
Following the proof of Lemma 2 in \cite{Kuba_Chen_Wen_Wen_Sun_Wang_Yang_2021}, we know that HAPPO has the same monotonic improvement bound as that of CoPPO. For the monotonic improvement of a single agent, we formulate the surrogate objective of agent $i$ using HAPPO as $\mathcal{J}(\hat{\boldsymbol{\pi}}^{i-1})+$ $\frac1{1-\gamma}\mathbb{E}_{(s,\boldsymbol{a})\sim(d^\pi,\hat{\boldsymbol{\pi}}^i)}[A^{\boldsymbol{\pi}}(s,\boldsymbol{a})]-\frac1{1-\gamma}\mathbb{E}_{(s,\boldsymbol{a})\sim(d^{\boldsymbol{\pi}},\hat{\boldsymbol{\pi}}^{i-1})}[A^{\boldsymbol{\pi}}(s,\boldsymbol{a})]$.
$$
\begin{aligned}
&\left|\mathcal{J}(\hat{\boldsymbol{\pi}}^{i})-\mathcal{J}(\hat{\boldsymbol{\pi}}^{i-1})-\frac{1}{1-\gamma}\mathbb{E}_{(s,\boldsymbol{a})\sim(d^{\boldsymbol{\pi}},\hat{\boldsymbol{\pi}}^{i})}[A^{\boldsymbol{\pi}}]+\frac{1}{1-\gamma}\mathbb{E}_{(s,\boldsymbol{a})\sim(d^{\boldsymbol{\pi}},\hat{\boldsymbol{\pi}}^{i-1})}[A^{\boldsymbol{\pi}}(s,\boldsymbol{a})]\right| \\
\leq&\frac{1}{1-\gamma}\left|\mathbb{E}_{(s,\boldsymbol{a})\sim(d^{\hat{\boldsymbol{\pi}}^i},\hat{\boldsymbol{\pi}}^i)}\left[A^{\hat{\boldsymbol{\pi}}^{i-1}}\right]-\mathbb{E}_{(s,\boldsymbol{a})\sim(d^{\boldsymbol{\pi}},\hat{\boldsymbol{\pi}}^i)}\left[A^{\boldsymbol{\pi}}\right]\right|+\frac{1}{1-\gamma}\left|\mathbb{E}_{(s,\boldsymbol{a})\sim(d^{\boldsymbol{\pi}},\hat{\boldsymbol{\pi}}^{i-1})}[A^{\boldsymbol{\pi}}(s,\boldsymbol{a})\right.  \\
\leq&\frac{1}{1-\gamma}\left|\mathbb{E}_{(s,\boldsymbol{a})\sim(d^{\boldsymbol{\pi}^{i}},\hat{\boldsymbol{\pi}}^{i})}\left[A^{\hat{\boldsymbol{\pi}}^{i-1}}\right]-\frac{1}{1-\gamma}\mathbb{E}_{(s,\boldsymbol{a})\sim(d^{\boldsymbol{\pi}},\hat{\boldsymbol{\pi}}^{i})}\left[A^{\hat{\boldsymbol{\pi}}^{i-1}}\right]\right| \\
&+\frac{1}{1-\gamma}\left|\mathbb{E}_{(s,\boldsymbol{a})\sim(d^{\boldsymbol{\pi}},\hat{\boldsymbol{\pi}}^{i})}\left[A^{\hat{\boldsymbol{\pi}}^{i-1}}\right]-\mathbb{E}_{(s,\boldsymbol{a})\sim(d^{\boldsymbol{\pi}},\hat{\boldsymbol{\pi}}^{i})}\left[A^{\boldsymbol{\pi}}\right]\right|+2\frac{1}{1-\gamma}\sum_{j\in e^{i}}\alpha^{j}e^{\boldsymbol{\pi}} \\
\leq&4\epsilon^{\hat{\boldsymbol{\pi}}^{i-1}}\alpha^{i}\sum_{t=0}^{\infty}\gamma^{t}(1-(1-\sum_{j\in(e^{i}\cup\{i\})}\alpha^{j})^{t}) \\
&+\frac{1}{1-\gamma}\mathbb{E}_{(s,\boldsymbol{a})\sim(d^{\boldsymbol{\pi}},\boldsymbol{\pi}^{i})}\left[\left|A^{\hat{\boldsymbol{\pi}}^{i-1}}-A^{\boldsymbol{\pi}}\right|\right]+2\frac{1}{1-\gamma}\sum_{j\in e^{i}}\alpha^{j}e^{\boldsymbol{\pi}} \\
\leq&4\epsilon^{\hat{\boldsymbol{\pi}}^{i-1}}\alpha^{i}(\frac{1}{1-\gamma}-\frac{1}{1-\gamma(1-\sum_{j\in(e^{i}\cup\{i\})}\alpha^{j})})+\frac{1}{1-\gamma}\left[4\alpha^{i}\epsilon^{\hat{\boldsymbol{\pi}}^{i-1}}+4\sum_{j\in\boldsymbol{e}^{i}}\alpha^{j}\epsilon^{\boldsymbol{\pi}}\right]
\end{aligned}
$$

There exists uncontrollable term $\sum_{j\in(e^{i}\cup\{i\})}\alpha^{j}$ for agent $i$, which means HAPPO does not provide a guarantee for improving the expected performance $\mathcal{J}(\hat{\boldsymbol{\pi}}^{i})$ even if the total variation distances of consecutive policies are well constrained.
\begin{table}[b]
\caption{The notations and symbols used in this paper.}
\vskip 0.15in
\begin{center}
\begin{small}
\begin{tabular}{ll}
\toprule
Notation & Definition \\
\midrule
$S$    & The state space\\
$I$    & The set of agents\\
$n$    & The number of agents\\
$i$    & The agent index\\
$\boldsymbol{A}$    & The action space of agents\\
$A^i$    & The action space of agent $i$\\
$B$   & A collection of $m$ batches\\
$b_k$   &  The set of agents for the $k$th batch policy update\\
$e^{i}$   & The set of preceding agents updated before agents $i$\\
$B_k$   & The set of preceding agents updated before agents in batch $b_k$\\
$\Omega$  & The observation function\\
$R$      & The reward function\\
$P$      & The transition function\\
$\gamma$   & The discount factor\\
$t$   & The time-step $t$\\
$s_t$   & The state at time-step $t$\\
$o_t$  & The observation of agents at time-step $t$\\
$o_t^i$  & The observation of agent $i$ at time-step $t$\\
$a_t^i$   & The action of agent $i$ at time-step $t$\\
$\boldsymbol{a}_t$   & The joint action at time-step $t$\\
$r_t$    & The shared rewards of agents at time-step $t$\\
$r_{p:q}$    & The discounted cumulative rewards of agents from time-step $p$ to $q$, where $r_{p:q} = \sum_{l=0}^{q-p}(\gamma)^{l}R(s_{p+l},\boldsymbol{a}_{p+l})$\\
$Pr(\cdot|\pi)$   & The state probability function under $\pi$\\
$d^{\boldsymbol{\pi}}$   & The discounted state visitation distribution, where $d^\pi(s) = (1-\gamma)\sum_{t=0}^{\infty}\gamma^tPr(s_t= s|\pi)$\\

$V$   & The value function\\
$A$   & The advantage function \\
$\tau_t^i$   & The trajectory of agent $i$ at time-step $t$\\
$\mathbb{T}$   & The trajectory space $\mathbb{T} \equiv (\Omega \times \boldsymbol{A})^{*}$\\
$\pi^i$   & The policy of agent $i$\\
$\hat{\pi}^i$   & The updated policy of agent $i$\\
$\hat{\boldsymbol{\pi}}^i$   & The joint policy after updating agent i\\
$\pi^{b_k}$   & The joint policy of agents in batch $b_k$\\
$\hat{\pi}^{b_k}$   & The updated joint policy of agents in batch $b_k$\\
$\hat{\boldsymbol{\pi}}^{b_k}$   & The joint policy after updating batch $b_k$\\

\bottomrule
\end{tabular}
\end{small}
\end{center}
\vskip -0.1in
\end{table}

\begin{table}[ht]
\caption{The notations and symbols used in this paper.}
\vskip 0.15in
\begin{center}
\begin{small}
\begin{tabular}{ll}
\toprule
Notation & Definition \\
\midrule
$\boldsymbol{\pi}$   & The joint policy\\
$\hat{\boldsymbol{\pi}}$   & The updated joint policy\\

$A^{\boldsymbol{\pi}}$   & The advantage of joint policy $\boldsymbol{\pi}$\\
$A^{\boldsymbol{\pi},\hat{\boldsymbol{\pi}}^{b_{k-1}}}$ & The appropriation of the advantage $A^{\hat{\boldsymbol{\pi}}^{b_{k-1}}}$ utilizing samples  collected under $\boldsymbol{\pi}$ \\
$\delta_{t}$ & The temporal difference for $V(s_t)$, $\delta_{t} = r_t+\gamma V(s_{t+1})-V(s_t)$\\
$\delta_{t}^{dag}$ & The temporal difference of upper-level network for $V(s_t)$, $\delta^{dag}_{t} = r_{t:t+T} + \gamma V_{\phi}^{dag}(s_{t+T}) - V_{\phi}^{dag}(s_{t})$\\
$n_t$ & The future maximum time-steps.\\
$\lambda$ & The parameter controlling the bias
and variance.\\
$J(\boldsymbol{\pi})$   & The expected return / performance of the joint policy $\boldsymbol{\pi}$\\
$J(\boldsymbol{\hat{\boldsymbol{\pi}}^{b_k}})$   & The expected return / performance of the joint policy $\hat{\boldsymbol{\pi}}^{b_k}$ after updating batch $b_k$\\
$L_{\hat{\boldsymbol{\pi}}^{i-1}}(\hat{\boldsymbol{\pi}}^{i})$   & The surrogate objective of agent i\\
$L_{\hat{\boldsymbol{\pi}}^{b_{k-1}}}(\hat{\boldsymbol{\pi}}^{b_k})$   & The surrogate objective of agents in batch $b_k$\\
$G_{\boldsymbol{\pi}}(\hat{\boldsymbol{\pi}})$   & The surrogate objective of all agents\\
$\epsilon$   & The upper bound of an advantage function $\epsilon=\max_{b_k}\epsilon^{b_k}$\\
$\epsilon^{b_k}$   & The upper bound of the advantage function of policy $\hat{\boldsymbol{\pi}}^{b_{k-1}}$ before updating batch $b_k$\\
$D_{TV}$   & The total variation(TV) distance function\\
$D_{KL}$   & The Kullback-Leibler divergence function\\
$\alpha$   & The total variation distance between 2 policies\\
$\alpha^{b_k}$   & The maximum total variation distance between $\pi^{b_k}$ and $\hat{\pi}^{b_k}$  \\
$\xi^{b_k}$   & The off-policy correction error of policy $\boldsymbol{\hat{\pi}^{b_{k-1}}}$, $\xi^{b_k}=$
    $\max_{s,\boldsymbol{a}}|A^{\boldsymbol{\pi},\hat{\boldsymbol{\pi}}^{b_{k-1}}}(s,\boldsymbol{a})-A^{\hat{\boldsymbol{\pi}}^{b_{k-1}}}(s,\boldsymbol{a})|$\\
$\boldsymbol{\pi}_{ind}$ & The joint policy acquired by simultaneous policy optimization scheme\\
$\pi_{ind}^{i}$ & The policy of agent $i$ acquired by simultaneous policy optimization scheme \\
$\boldsymbol{\pi}$ & The joint policy produced by B2MAPO scheme\\
$\pi^{i}$ & The policy of agent $i$ produced by B2MAPO scheme\\
$V_{\boldsymbol{\pi}}$ & The value function of the joint batch-dependent policy \\
$V_{\boldsymbol{\pi}_{ind}}$ & The value function of the joint independent policy \\

$Q_{\boldsymbol{\pi}}$ & The action-value function of the joint batch-dependent policy \\
$Q_{\boldsymbol{\pi}_{ind}}$ & The action-value function of the joint independent policy \\
$G$ & The spinning subgraph DAG of the dependencies between different agents, $G=(V, E)$\\
$G_T$ & The estimated DAG graph matrix of the dependencies between different agents of period $T$\\
$G^{i}_T$ & The estimated DAG graph matrix of the dependencies between agent $i$ and other different agents of period $T$\\
$Z_T$ & The abstract representation of the current relationships between agents of period $T$\\
$Sr$ &The important sample rate between the current DAG policy and the old DAG policy, where $Sr(\theta) = \frac{\pi^{dag}_{\theta}(a_t | s_t)}{\pi^{dag}_{\theta_{old}}(a_t | s_t)}$\\
$\pi^{dag}_{\theta}$ & The DAG policy parameterized by $\theta$ \\

$V_{\phi}^{dag}$ &The estimate value function of DAG policy by critic network, which is parameterized by $\phi$\\

$ \hat{A}^{dag}$ &The estimate advantage function of DAG policy for the future $T$ steps\\
\bottomrule
\end{tabular}
\end{small}
\end{center}
\end{table}

\end{document}